\documentclass[a4paper,11pt]{article}
\usepackage[T1]{fontenc}
\usepackage[utf8]{inputenc}
\usepackage{jheppub}
\usepackage{orcidlink}
\usepackage{slashed}
\usepackage{bbold}
\usepackage{multirow}
\usepackage{graphicx}
\usepackage{subcaption}
\usepackage{feynmp-auto}
\newcommand{\Rangle}{\ensuremath{r}}
\newcommand{\RPhaseInput}{\ensuremath{{\omega_{22}}}}
\newcommand{\RPhaseReplaced}{\ensuremath{{\omega_{32}}}}
\newcommand{\RPhaseOverall}{\ensuremath{{\phi_R}}}

\newcommand{\br}{\ensuremath{{\operatorname{BR}}}}
\newcommand{\rate}{\ensuremath{{\operatorname{CR}}}}

\newcommand{\im}{\ensuremath{{\operatorname{Im}}}}
\newcommand{\sign}{\ensuremath{{\operatorname{sign}}}}
\newcommand{\plane}{\ensuremath{{\RPhaseInput-\Rangle\text{ plane}}}}

\newcommand{\positiveSolution}{{\ensuremath{\RPhaseReplaced_+}}}
\newcommand{\negativeSolution}{{\ensuremath{\RPhaseReplaced_-}}}

\newcommand{\photonFactor}{{\ensuremath{|\Lambda|m_{H^\pm}^2}}}
\newcommand{\boxFactor}{{\ensuremath{\Lambda^2m_{H^\pm}^2}}}

\newcommand{\code}[1]{\texttt{#1}}
\newcommand{\indexvalign}{\ensuremath{^{\vphantom{i}}}}
\newcommand{\verticalequal}{\rotatebox{90}{$=$}}
\newcommand{\xequal}[1]{\stackrel{#1}{=}}
\newcommand{\hc}{\ensuremath{h.c.}}
\newcommand{\pole}[1]{\ensuremath{m_{#1}^{\text{pole}}}}

\newcommand{\lhs}{lhs.}
\newcommand{\rhs}{rhs.}
\newcommand{\norm}[1]{|#1|^2}

\newcommand{\sfrac}[2]{{\textstyle \frac{#1}{#2}}}
\newcommand{\nuratio}{{\ensuremath{t_{32}}}}

\newcommand{\mueg}{\ensuremath{{\mu\to e\gamma}}}
\newcommand{\taueg}{\ensuremath{{\tau\to e\gamma}}}
\newcommand{\taumug}{\ensuremath{{\tau\to \mu\gamma}}}
\newcommand{\mueee}{\ensuremath{{\mu\to3e}}}
\newcommand{\taueee}{\ensuremath{{\tau\to 3e}}}
\newcommand{\taummm}{\ensuremath{{\tau\to 3\mu}}}
\newcommand{\taumee}{\ensuremath{{\tau\to \mu ee}}}
\newcommand{\tauemm}{\ensuremath{{\tau\to e\mu\mu}}}
\newcommand{\muec}{\ensuremath{{\mu\to e\text{ conversion}}}}

\newcommand{\flavorY}[2]{\ensuremath{{Y^{(#1)}_{#2}}}}
\newcommand{\niceY}[2]{\ensuremath{{Y^{(#1\prime)}_{#2}}}}
\newcommand{\massY}[2]{\ensuremath{{Y^{(#1\prime\prime)}_{#2}}}}
\newcommand{\smYukawa}[2]{\ensuremath{{Y^{#1}_{#2}}}}

\newcommand{\flavorNu}[2]{\ensuremath{{\nu^{#1}_{#2}}}}
\newcommand{\niceNu}[2]{\ensuremath{{\nu^{\prime #1}_{#2}}}}
\newcommand{\massNu}[2]{\ensuremath{{\nu^{\prime\prime #1}_{#2}}}}

\newcommand{\oneLoopNuMatrix}{\ensuremath{{\hat\Sigma}}}

\newcommand{\fullRotation}[1]{%
	\ifnum#1=3\ensuremath{{U}}%
	\else\ifnum#1=4\ensuremath{{\tilde{U}}}%
	\fi\fi
}

\newcommand{\niceRotation}[1]{%
	\ifnum#1=3\ensuremath{{V}}%
		\else\ifnum#1=4\ensuremath{{\tilde{V}}}%
	\fi\fi
}

\newcommand{\seesawRotation}[1]{%
	\ifnum#1=3\ensuremath{{S}}%
		\else\ifnum#1=4\ensuremath{{\tilde{S}}}%
	\fi\fi
}

\newcommand{\loopRotation}[1]{%
	\ifnum#1=2\ensuremath{{\hat{R}}}%
		\else\ifnum#1=3\ensuremath{{R}}%
			\else\ifnum#1=4\ensuremath{{\tilde{R}}}%
	\fi\fi\fi
}

\newcommand{\pmns}[1]{{\ensuremath{U^{#1}_{\text{\gls{pmns}}}}}}
\usepackage{glossaries-extra}
\usepackage{caption}

\setabbreviationstyle[acronym]{long-short}

\glssetcategoryattribute{acronym}{nohyperfirst}{true}

\newacronym{dm}{DM}{Dark matter}
\newacronym{idm}{IDM}{Inert Doublet model}
\newacronym{sm}{SM}{Standard model}
\newacronym{bsm}{BSM}{beyond Standard model}
\newacronym{gnm}{GNM}{Grimus-Neufeld model}
\newacronym{thdm}{2HDM}{two-Higgs-doublet model}
\newacronym{vev}{VEV}{vacuum expectation value}
\newacronym{pmns}{PMNS}{Pontecorvo-Maki-Nakagawa-Sakata}
\newacronym{nh}{NO}{Normal ordering}   
\newacronym{ih}{IO}{Inverted ordering} 
\newacronym{gim}{GIM}{Glashow-Iliopoulos-Maiani}
\newacronym{clfv}{cLFV}{Charged Lepton Flavour Violation}

\newacronym{no1}{NO-1}{realistic benchmark point}
\newacronym{no2}{NO-2}{intermediate benchmark point}
\newacronym{no3}{NO-3}{hopeless benchmark point}
\title{\boldmath Charged lepton flavor violating processes in the Grimus-Neufeld model}

\author[a,1]{Vytautas D\=ud\.enas\,\orcidlink{0000-0001-9405-9959},\note{Corresponding author.}}
\author[a]{Thomas Gajdosik\,\orcidlink{0000-0002-4355-8878},}
\author[b]{Uladzimir Khasianevich\,\orcidlink{0000-0003-0255-0674},}
\author[c]{Wojciech Kotlarski\,\orcidlink{0000-0002-1191-6343},}
\author[b]{Dominik St\"ockinger}

\affiliation[a]{
Institute of Theoretical Physics and Astronomy, Faculty of Physics, Vilnius University,
9 Sault\.ekio, LT-10222 Vilnius, Lithuania%
}

\affiliation[b]{
Institut f\"ur Kern- und Teilchenphysik,
TU Dresden, Zellescher Weg 19, 01069 Dresden, Germany
}

\affiliation[c]{
National Centre for Nuclear Research, Pasteura 7, 02-093 Warsaw, Poland
}

\emailAdd{vytautasdudenas@inbox.lt}
\emailAdd{thomas.gajdosik@cern.ch}
\emailAdd{uladzimir.khasianevich@tu-dresden.de}
\emailAdd{wojciech.kotlarski@ncbj.gov.pl}
\emailAdd{dominik.stoeckinger@tu-dresden.de}

\abstract{%
Charged Lepton Flavour Violating~(cLFV) decays constrain 
the relationship between the neutrino and the scalar sectors of
the Grimus-Neufeld model~(GNM), an appealing minimal model of neutrino masses.
It turns out, that in the scenario, where the seesaw scale is lower than the electroweak one, cLFV is completely defined by the new Yukawa interactions between the additional single heavy Majorana neutrino, the second Higgs doublet and the lepton doublets.
Therefore, we derive a useful parameterization for the Yukawa couplings which reproduces by construction the correct PMNS matrix and the correct neutrino masses for both Normal and Inverted ordering at one-loop level. 
We embed this scenario in the \texttt{FlexibleSUSY} spectrum-generator generator to perform parameter scans.
Focusing on the tiny seesaw scale, we show that current \ensuremath{\mu\to e\gamma}
limits provide significant constraints on the scalar sector, and we
evaluate the impact of future cLFV $\tau$-decay searches for the cases
of discovery or non-discovery. 
The tiny seesaw scale makes the neutrino sector and the cLFV processes in the GNM similar to the scotogenic and the scoto-seesaw models, so we provide constraints for these models as well.%
}

\begin{document}
\maketitle
\flushbottom
\section{Introduction}

Many years have passed since the discovery of neutrino oscillations~\cite{Athanassopoulos:1997pv,Fukuda:1998mi,Eguchi:2002dm,Ahmad:2002jz}, yet massive neutrinos are still not in the \gls{sm}.
That is not surprising: it is extremely hard to either confirm or exclude all the possible mechanisms that generate neutrino masses due to their weak impact on the sectors that we can actually see in the experiments,
see e.g. studies of the low energy effects of the seesaw mechanisms in~\cite{Abada:2007ux, Coy:2021hyr}. 
However, having in mind the impressive precision of current \gls{clfv} experiments~\cite{MEG:2020zxk, BaBar:2009hkt,  Hayasaka:2010np, SINDRUM:1987nra} 
some of the scenarios can actually lead to possible signatures and/or restrictions on the parameter spaces, and will become even more restricting in the future, given planned experiments \cite{MEGII:2018kmf,Belle-II:2018jsg,Wasili:2020ksf,COMET:2018auw,Cirigliano:2021img}. Hence, it makes sense to look at the most constraining scenarios to narrow down the possible space of the neutrino mass mechanisms. 

The simplest neutrino mass mechanisms can be put into three categories: 
inducing neutrino masses at tree-level (e.g. all types of seesaw \cite{Minkowski:1977sc, Yanagida:1979as, Mohapatra:1979ia}), 
generating them at loops (see e.g. \cite{Zee:1980ai, Pilaftsis:1991ug, Ma:2006km}), 
and combining these mechanisms (e.g. \cite{Grimus:1989pu, MANDAL2021136458}).
These mass mechanisms can be embedded into more exotic models, for example, the seesaw mechanism can arise from extra dimensions \cite{Girmohanta:2020llh, Girmohanta:2021gpf}, or radiative neutrino generation can be realized with composite Higgses \cite{Cacciapaglia:2020psm, Rosenlyst:2021tdr}.   
While virtually all scenarios can fit the current experimental
constraints, it is far harder to claim the strictly excluded regions unambiguously, especially when one studies more general scenarios with an overwhelming number of free parameters (see e.g.  \cite{Jurciukonis:2021izn}). 
A look at less general models, with more exposed and isolated effects on the \gls{clfv} from the neutrino sector, naturally provides a solution to this problem. 

The scotogenic model~\cite{Ma:2006km} is probably the most popular one in this respect, partially due to a manageable amount of free parameters.
With  an imposed $Z_2$ symmetry, it has the scalar sector of the inert doublet model~(IDM)~\cite{PhysRevD.18.2574}. 
The \gls{clfv} in the scotogenic model comes purely from the radiative contributions of heavy neutrinos and scalar dark matter candidates~\cite{Toma:2013zsa,  Vicente:2014wga}.
Thus, the neutrino sector acts as a bridge between the dark scalar sector and \gls{clfv}.

There is an interesting variation, called the scoto-seesaw model \cite{Rojas:2018wym, Mandal:2021yph}, with even fewer degrees of freedom introduced. 
In both models, scotogenic and scoto-seesaw, one obtains larger contributions to the \gls{clfv} for lower heavy neutrino masses.
This potentially implies stronger restrictions on the scalar sector. 
We define a \emph{tiny} seesaw scale to be lower than the electroweak scale (studied in e.g.~\cite{Dinh:2021fmt, Klaric:2020phc,Rasmussen:2016njh,Bondarenko:2018ptm,Drewes:2016jae}), to discriminate between the low seesaw scale that is sometimes considered to include the TeV scale (e.g.~\cite{Chen:2011de,Lopez-Pavon:2015cga,Dinh:2013vya}). 
Then, in the tiny seesaw scale, the scotogenic and the scoto-seesaw
models give essentially the same contribution to \gls{clfv} as we get
in the \gls{gnm} \cite{Grimus:1989pu} \textemdash{} a seesaw extended \gls{thdm}. 
This connection is caused by an approximate $Z_2$ symmetry in the Yukawa sector~\cite{Dudenas:2022qcw} (discussed in the next section), which makes the radiative neutrino mass generation in the \gls{gnm} similar to the one in the scotogenic and scoto-seesaw models. 

Global symmetries, such as $Z_2$, are extremely useful for classifying the scenarios of \gls{bsm} physics. 
Yet they are not something fundamental. 
If one imposes a global symmetry only on one part of the Lagrangian, while breaking at some other part of it,
at some loop order it eventually leads to breakage of the imposed symmetry in the sector of interest too.
From one point of view, one might conclude that imposing a global symmetry is only consistent if it is done on the full Lagrangian. 
On the other hand, if one looks at global symmetries as something that just helps us to classify possible scenarios of \gls{bsm} physics, there is no real need to insist on satisfying it exactly at all loop levels. 
For example, the CP-symmetric \gls{thdm} potential might get CP-violating corrections from the quark sector, as discussed in \cite{Fontes:2021znm}.
There are many studies on CP-conserving \gls{thdm} potential, which nevertheless are meaningful and explore an important possible scenario of the scalar sector. 
This naturally leads to a question: what if some parameters in the Lagrangian are so small, that we can \emph{almost} see the global symmetry?
This generalizes the concept of the global symmetry of the Lagrangian by allowing small deviations from zero of the symmetry breaking terms. 
Following this logic, we will loosely define the approximate symmetry to be the case in which the Lagrangian parameters that explicitly break the global symmetry are not necessarily zero, but are very small (in our case, smaller than $O(10^{-7})$ ). 
The \gls{gnm} satisfies this requirement in the tiny seesaw limit. 

By itself, the \gls{gnm} is an appealing model of neutrino masses
which postulates the existence of only a single heavy neutrino
to accommodate neutrino masses and mixings, while the other models require at least two additional fields.
The \gls{gnm} is then particularly attractive in the tiny seesaw parameter region: it has less particles and, because of the approximate (but not exact) $Z_2$ symmetry, it can express the main phenomenological features of both the scotogenic and the scoto-seesaw models of the \gls{clfv} processes and the neutrino sector.
Investigating this parameter region we will address the following question in this paper: Which restrictions are imposed by the \gls{clfv} on the scalar sector in the \gls{gnm}?

In section~\ref{sec:GN model}, we give an overview of the \gls{gnm} by introducing the scalar and Yukawa sectors. 
Then we present the one-loop neutrino mass calculation.
It leads to a parameterization of the flavor basis Yukawa couplings, similar to the Casas-Ibarra parameterization~\cite{Casas:2001sr} that automatically reproduces neutrino masses and mixings at one-loop level. 
We close the section by noting the existing special parameter points in this parameterization, the checks, the numerical stability, and the limitations of our study, using \code{FlexibleSUSY}~\cite{Athron:2014yba, Athron:2017fvs, Athron:2021kve}. 
In section~\ref{sec:clfv processes} we present the analytical expressions for \gls{clfv} processes. 
In section~\ref{sec:strategy} we give a short recap of the most important parameters from the Yukawa sector and the scalar sector that control the branching ratios of \gls{clfv} processes and we lay out the strategy of our phenomenological study. 
We then pursue this strategy in section~\ref{sec:pheno}, give an interpretation of the results, and conclude our study in section~\ref{sec:conclusions}. 
Technical details, such as the explicit derivation of the parametrization, peculiarities of the minimal free parameter set in the Yukawa sector, and the numerical values of some special cases can be found in the appendices.

\section{The Grimus-Neufeld model}\label{sec:GN model}
The \gls{gnm} is a general \gls{thdm} extended with one
single sterile neutrino.
In the limit of the \emph{tiny seesaw scale} discussed below, the Yukawa sector is 
approximately $Z_2$ symmetric~\cite{Dudenas:2022qcw}, thus its predictions for \gls{clfv}
become similar to the scotogenic and scoto-seesaw models. 
Scotogenic and the scoto-seesaw models also predict scalar \gls{dm}, which is a direct consequence of the $Z_2$ symmetry. 
The approximate $Z_2$ symmetry in the \gls{gnm} will make the potential scalar \gls{dm} candidate in the \gls{gnm} not as stable.
Yet to be sure if \gls{gnm} has a viable \gls{dm} candidate in this scenario, a dedicated analysis is needed, which is beyond the scope of this work.  
We do not discuss \gls{dm} in this paper.

\subsection{Scalar and Yukawa sectors}

In principle, \gls{clfv} in the \gls{gnm} can be analyzed for
a general \gls{thdm} scalar sector.
However, to highlight the similarities between models we consider the scalar potential to be $Z_2$ symmetric. 
The Higgs sector of the model contains two Higgs doublets
$H_{1,2}$. The potential takes the form:
\begin{equation}
\newcommand{\Higgs}[1]{H_{#1}^{\vphantom{\dagger}}}
\newcommand{\CHiggs}[1]{H_{#1}^\dagger}
\begin{aligned}
V& =
	m_{11}^{2\vphantom{\dagger}}\CHiggs1\Higgs1
	+\frac{\lambda_{1}}{2}(\CHiggs1\Higgs1)^{2}
	+m_{22}^{2}\CHiggs2\Higgs2
	+\frac{\lambda_{2}}{2}(\CHiggs2\Higgs2)^{2} \\
&
	+\lambda_{3}(\CHiggs1\Higgs1)(\CHiggs2\Higgs2)
	+\lambda_{4}(\CHiggs2\Higgs1)(\CHiggs1\Higgs2)
	+\Big[
		\frac{\lambda_{5}}{2}(\CHiggs2\Higgs1)^2 + \hc
	 \Big]
.
\end{aligned}
\label{eq:Higgs potential}
\end{equation}
Only the first Higgs doublet $H_{1}$
acquires a \gls{vev}{} $v$, and we parametrize the two doublets as:
\begin{equation}
H_{1}=\left(\begin{matrix}
G_{W}^{+}\\
\frac{1}{\sqrt{2}}\left(v+h+iG_{Z}\right)
\end{matrix}\right),
\quad
H_{2}=\left(\begin{matrix}
H^{+}\\
\frac{1}{\sqrt{2}}\left(H+iA\right)
\end{matrix}\right)
.
\label{eq:Higgs basis}
\end{equation}

All the \gls{sm} particles, including $H_1$, are assigned an even parity under the $Z_2$ symmetry, while $H_{2}$ and the additional sterile neutrino $N$ are odd.\footnote{For fermions we use the notation of two component Weyl spinors.}
The latter enters the Lagrangian with, in general, complex Majorana mass term $M$
and new Yukawa-like coupling  \flavorY{i}{j} to the $SU(2)$ lepton
doublets $\ell_j$, where $j$ denotes the generation:
\begin{equation}
\mathcal{L} \ni 
-\frac{1}{2}MNN
-\flavorY{i}{j} \ell_j\indexvalign\epsilon H_i\indexvalign N+\hc
\label{eq:lepton neutrino yukawa}
\end{equation}
The matrix $\epsilon=i\sigma_2$ combines the two doublets to an $SU(2)$ invariant product. 
The Majorana mass $M$ is made real by adjusting the phase of $N$.
Terms with the neutrino Yukawa couplings \flavorY{1}{} to the first Higgs doublet in eq.~\eqref{eq:lepton neutrino yukawa} explicitly break the $Z_2$ symmetry.
The \gls{sm}-like $Z_2$-preserving Yukawa sector reads as:
\begin{equation}
\mathcal{L} \ni
-\smYukawa{}{kj} \tilde{H}_1 \epsilon \ell_j\indexvalign e^c_k+\hc
\label{eq:lepton neutrino yukawa-2}
\end{equation}
Note the usage of $\widetilde{H}_1 = \epsilon H^*_1$ allowing for another $SU(2)$ invariant product with opposite electric charge.
In the flavor basis charged lepton masses are defined by $\smYukawa{}{ff}=\sqrt{2}m_{f}/v$, $f=e,\mu,\tau$, which is also the mass eigenstate basis for charged leptons, but not the mass eigenstate basis for neutrinos. 

When the Higgs acquires a \gls{vev}, the Lagrangian in eq.~\eqref{eq:lepton neutrino yukawa} leads to the two non-vanishing Majorana masses for neutrinos, light $m_3$ and heavy $m_4$ (in mass eigenstate basis):
\begin{equation}
\mathcal{L} \ni -\frac{1}{2} m_3 \niceNu{}{3} \niceNu{}{3} - \frac{1}{2} m_4 \niceNu{}{4} \niceNu{}{4}\,,
\end{equation} 
where
\begin{equation}
y^{2}:=  \sum_{i} \big|\flavorY{1}{i}\big|^2=\frac{2m_{3}m_{4}}{v^{2}}
\,,\quad
M=m_{4}-m_{3}
\,,
\label{eq:seesaw relations}
\end{equation}
which is the usual type-I seesaw mechanism relating the Yukawa coupling \flavorY{1}{i}, the light neutrino mass $m_3$, and the seesaw scale $m_4\gg m_3$.
The $m_3$ mass is associated with a light neutrino mass and is of $O(0.01\ \text{eV})$.
If $y\to 0 $ for fixed $m_4$ then ${m_3 \to 0}$, and the seesaw mechanism is ``turned off'' for the exact $Z_2$ symmetry. 
For $y\neq 0$, the seesaw mechanism applies, and for fixed $m_3$, eq.~\eqref{eq:seesaw relations} can be treated as a relationship between the seesaw scale and the $Z_2$ breaking parameter, allowing us to define an approximate $Z_2$ symmetry in the \gls{gnm} by the inequalities:
\begin{equation}
m_4 \lesssim 10\ \text{GeV}
\quad \Leftrightarrow \quad
y \lesssim 10^{-7}
.
\label{eq:seesaw-limit}
\end{equation}
The restriction of eq.~\eqref{eq:seesaw-limit} ensures that the $Z_2$-breaking Yukawa coupling is at least an order of magnitude smaller than the Yukawa coupling of the electron.

There are at least two non-vanishing masses for light neutrinos. In the \gls{gnm} the mass of the other \gls{sm} neutrino is generated at one-loop level via interactions with $H_2$.
Motivated by a broken Peccei-Quinn symmetry, small $\lambda_{5}$ values successfully generate radiative neutrino masses with relatively large Yukawa couplings to the second Higgs doublet, namely \flavorY{2}{}.
Just as the limit $y \to 0$ turns off the seesaw mechanism, setting
$\lambda_{5} \to 0$ turns off the radiative mass generation leaving 
one of the needed neutrino masses unexplained.
In the \gls{gnm} the lightest neutrino is massless. 

The $Z_2$ symmetry in the Yukawa sector leads to the similarity between the \gls{gnm}, the scotogenic, and the scoto-seesaw models
in the \gls{clfv} phenomenology, even though the Yukawa sectors are slightly different. 
The scotogenic model~\cite{Ma:2006km} has an exact $Z_2$ symmetry, forcing $y\to0$ and turning off the seesaw mechanism at the cost of adding 2 additional heavy, $Z_2$-odd neutrino states. 
Then the Yukawa couplings \flavorY{2}{}, which lead to radiative mass generation, are contained in a $3 \times 3$ matrix instead of a $3\times 1$ as in the \gls{gnm}. 

In the scoto-seesaw model~\cite{Rojas:2018wym, Mandal:2021yph}, one has an exact $Z_2$ symmetry turning off the seesaw mechanism for the $Z_2$-odd $N$, at the cost of adding one $Z_2$-even sterile neutrino, for which, only the seesaw mechanism is allowed. 
This effectively gives one additional independent parameter, the mass of the $Z_2$-even neutrino in contrast to the \gls{gnm}. In turn, this allows to control the sizes of parameters that enter the radiative and the seesaw mass mechanisms independently from each other in the scoto-seesaw model, while they are related in \gls{gnm}. 

In the \gls{gnm}, the two \gls{sm}-like massive neutrino mass states mix in general. To get a convenient parameterization for the Yukawa couplings, one solves the equations for neutrino masses and mixings at one-loop level. 
We further describe the needed rotations, convenient basis and neutrino mass generation in \gls{gnm} leading to this parameterization in detail in the following sub-sections. 
\subsection{Rotations for neutrinos}
The \gls{gnm} contains four neutrino states, comprised of the three
neutrino components $\flavorNu{}{i}$ of the lepton doublets, and the
single sterile neutrino $N$. At tree level, neutrino masses arise from the
Majorana mass term $M$ and the Yukawa coupling vector $\flavorY{1}{}$,
coupling the neutrino states to the \gls{vev}.
These tree-level mass terms of eq.~\eqref{eq:lepton neutrino yukawa} give rise
to the $4\times4$ neutrino mass matrix $M_{\nu}^{F}$ in flavor basis
$\{\flavorNu{}{i}, N\}$ as shown with the first matrix in
figure~\ref{fig:neutrino-rotations}.  
To find the physical neutrino states and their masses, one has to
include one-loop corrections and diagonalize the  mass matrix by
unitary transformations.\footnote{Our conventions for unitary rotations are the ones
  of ref.~\cite{Haber:1984rc}, as implemented in \code{FlexibleSUSY}:
\begin{equation*}
\text{for}\
UU^\dagger = \mathbb{1}:
\quad
\psi^{old}_k = U_{ik}^* \psi^{new}_i
\,,\quad
m \xequal{\text{Takagi}} U^T m_{diag} U
\,,\quad
m \xequal{\text{SVD}} U_1^T m_{diag} U_2
\,.
\end{equation*}}
\begin{figure}[t!]
\newcommand{\order}[1]{\ensuremath{0^{#1\ell}}}
\centering
\begin{equation*}
\begin{matrix}
	\left(
	\begin{matrix}
	\order{0}_{3\times 3} & \frac{v}{\sqrt{2}} \flavorY{1}{} \\
	\frac{v}{\sqrt{2}} \flavorY{1}{}^T  & M
	\end{matrix}
	\right)
	&
	\xrightarrow{\niceRotation{4}}
	&
	\left(
	\begin{matrix}
	\order{1} & \order{1} & \order{1} & \order{1} \\
	\order{1} & \order{0} & \order{0} & \order{0} \\
	\order{1} & \order{0} & \order{0} & i\frac{vy}{\sqrt{2}} \\
	\order{1} & \order{0} & i\frac{vy}{\sqrt{2}} & M
	\end{matrix}
	\right)
	&
	\xrightarrow{\seesawRotation{4}}
	&
	\left(
	\begin{matrix}
	\order{1} & \order{1} & \order{1} & \order{1} \\
	\order{1} &  &  & \order{1} \\
	\order{1} & \multicolumn{2}{c}{\smash{\raisebox{.5\normalbaselineskip}{\oneLoopNuMatrix{}}}} & \order{1} \\
	\order{1} & \order{1} & \order{1} & m_4 + \order{1}
	\end{matrix}
	\right)
	&
	\xrightarrow{\loopRotation{4}}
	&
	\hat m
	\\
	\verticalequal & & & & & & \verticalequal
	\\
	M_\nu^F & & & & & &
	\fullRotation{4}^{*} M_\nu^F \fullRotation{4}^{\dagger}
	\\
	\flavorNu{}{\alpha} := \{\flavorNu{}{i}, N\} & &
	\niceNu{}{\alpha}
	& &
	\newlength{\dummyw}
	\settowidth{\dummyw}{$\approx\niceY{i}{}$\hfill}
	\parbox{\dummyw}{$\approx\niceNu{}{\alpha}$}
	 & &
	\massNu{}{\alpha}
	\\
	\flavorY{i}{} & &
	\niceY{i}{} & &
	\approx\niceY{i}{} & &
	\massY{i}{}
\end{matrix}
\end{equation*}
\caption{
	A sequence of rotations for neutrino mass matrix $M_\nu^F$.
	Explicit entries show quantities used in the paper. 
	We denote an expression that is zero exactly at tree-level by \order{0}. 
	With \order{1} we mean an approximated one-loop zero, where terms 
	proportional to $y^2 \times \text{loop}$ and $ \frac{m_3}{m_4} \times \text{loop}$ 
	are neglected. 
	The matrix \oneLoopNuMatrix{} contains non-vanishing expressions 
	for the processes $\niceNu{}{i}\rightarrow \niceNu{}{j}$ at one-loop 
	level, which are not further suppressed by $y^2$ or $\frac{m_3}{m_4}$.
}
\label{fig:neutrino-rotations}
\end{figure}

The mass matrix $M_{\nu}^{F}$ is diagonalized by a unitary matrix
\fullRotation{4} to a diagonal  matrix $\hat{m}$, see 
figure~\ref{fig:neutrino-rotations}. The diagonalization matrix
\fullRotation{4} can be decomposed into a product of three unitary
matrices: 
\begin{equation}
\fullRotation{4}^{*}M_{\nu}^{F}\fullRotation{4}^{\dagger}
:=
\loopRotation{4}^{*}
\seesawRotation{4}^{*}
\niceRotation{4}^{*}
M_{\nu}^{F}
\niceRotation{4}^{\dagger}
\seesawRotation{4}^{\dagger}
\loopRotation{4}^{\dagger}
=:
\hat{m}
\,.
\label{eq:all rotations}
\end{equation}
This decomposition is useful because each rotation leads to a separate physical consequence or highlights important details about the Yukawa couplings.
The sequence of these three rotations and their effect on the mass matrix is schematically shown in figure~\ref{fig:neutrino-rotations}.
In the following, we will discuss the diagonalization steps in more
detail. It will be useful to introduce the decomposition of the
relevant $4\times4$ matrices as:
\begin{equation}\label{eq:four-four-matrices}
\fullRotation{4} \approx\bigg(\begin{matrix}
\fullRotation{3} & \mathbb{0}\\
\mathbb{0} & \text{1}
\end{matrix}\bigg)
\,,\quad
\niceRotation{4}=\bigg(\begin{matrix}
\niceRotation{3} & \mathbb{0}\\
\mathbb{0} & \text{1}
\end{matrix}\bigg)
\,,\quad
\loopRotation{4}=\bigg(\begin{matrix}
\loopRotation{3} & \mathbb{0}\\
\mathbb{0} & \text{1}
\end{matrix}\bigg)
\,,\quad
\loopRotation{3}=\bigg(\begin{matrix}
1 & \mathbb{0}\\
\mathbb{0} & \loopRotation{2}
\end{matrix}\bigg)
\,,
\end{equation}
where \fullRotation{3}, \niceRotation{3}  and \loopRotation{3} are $3\times3$ unitary matrices and \loopRotation{2}
is a $2\times2$ unitary matrix.
We will define the matrix \seesawRotation{4} in eq.~\eqref{eq:definition_S4} and explain the approximate equality right after eq.~\eqref{eq:approximation}.

In the Lagrangian of eq.~\eqref{eq:lepton neutrino yukawa} new complex Yukawas \flavorY{i}{} are introduced, which carry 12 degrees of freedom in total.
As the first step of the diagonalization process we use the freedom to rotate flavor neutrinos \flavorNu{}{\alpha} to choose a convenient basis.
This is done by the matrix \niceRotation{3} in the following way:
\begin{equation}
\left.
\begin{alignedat}{3}
\flavorY{1}{j}\niceRotation{3}_{ij}^{*}
& =\{0,0,iy\}_{i}
&& =: \niceY{1}{i} \
&& \Leftrightarrow\ \text{5 conditions}
\,,\\
\flavorY{2}{j}\niceRotation{3}_{ij}^{*}
& =\{0,d,id^{\prime}\}_{i}
&& =: \niceY{2}{i} \
&& \Leftrightarrow\ \text{3 conditions}
\,.
\end{alignedat}
\right\} \quad\ \text{with}\ y\,,d>0\,,\ d^{\prime}\in\mathbb{C}\,,
\label{eq:GN parameters}
\end{equation}
i.e.\ the two Yukawa vectors $\flavorY{i}{}$ are transformed into two
positive and one complex parameter $y,d,d^{\prime}$. 
Thus the 8 degrees of freedom of the unitary matrix $V$ are determined by \lhs{} of eq.~\eqref{eq:GN parameters}, while the remaining 4 are shown in the \rhs{} of eq.~\eqref{eq:GN parameters}.
The only unfixed degree of freedom in a matrix $V$ is a single phase:
\begin{equation}
\niceRotation{3}\to \text{diag}(e^{i\alpha_{1}}\,,1\,,1)\niceRotation{3}\,.
\label{eq:phase freedom}
\end{equation}
This phase, $\alpha_1$, is used to absorb one of the Majorana phases $\eta_i$
of the \gls{pmns} matrix. 
Additionally, 3 degrees of freedom in $V$ can be absorbed into the phase definitions of neutrinos, leading to 5 physical degrees of freedom in $V$, which are later related to the 3 angles, 1 CP phase, and 1 Majorana phase of the \gls{pmns} matrix. 2 of the 4 degrees of freedom of the \rhs{} of eq.~\eqref{eq:GN parameters} will be related to neutrino masses. 

It is clear from eq.~\eqref{eq:GN parameters} that one of the neutrinos doesn't interact with the Higgs bosons after the rotation \niceRotation{4}, which leads to vanishing contributions to its mass at tree level and at one-loop level. These vanishing elements of the mass matrix are shown by $0^{1\ell}$ in the top and left outer entries of the matrices in figure~\ref{fig:neutrino-rotations}.

At the second step of  figure~\ref{fig:neutrino-rotations}, the
tree-level mass
matrix has a typical seesaw structure. Correspondingly, the second diagonalization matrix \seesawRotation{4} is a tree-level seesaw transformation, which yields
a diagonal mass matrix at tree level with two non-vanishing tree-level
mass eigenvalues
$m_{3}$ and $m_{4}$, related by eq.~\eqref{eq:seesaw relations}. 
The appropriate seesaw rotation \seesawRotation{4} can be written as:
\begin{equation}
\seesawRotation{4}:=\left(\begin{matrix}\mathbb{1} & \mathbb{0}\\
\mathbb{0} & \begin{matrix}c_{S} & is_{S}\\
is_{S} & c_{S}
\end{matrix}
\end{matrix}\right)
\,,\quad
\text{with}\ c_{S}=\sqrt{\frac{m_{4}}{m_{4}+m_{3}}}
\,,\quad s_{S}=\sqrt{\frac{m_{3}}{m_{4}+m_{3}}}
\,.
\label{eq:definition_S4}
\end{equation}
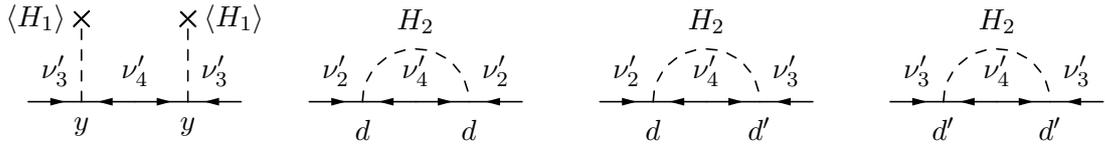
\begin{figure}
{
\unitlength = 1mm
\newcommand{\neutrinosed}[4]{
	\fmfframe(3,3)(3,5){
	\begin{fmfgraph*}(28, 11)
		\fmfpen{0.7}
		\fmfstraight
		\fmfset{arrow_len}{2mm}
		\fmfleft{bL,tL}
		\fmfright{bR,tR}
		\fmf{fermion,label=#1,label.side=left}{bL,vL}
		\fmf{fermion}{bM,vL}
		\fmf{fermion}{bM,vR}
		\fmf{fermion,label=#2,label.side=right}{bR,vR}
		\fmffreeze
		\fmf{dashes,left=1,label=$H_2$}{vL,vR}
		\fmfv{label=\niceNu{}{4},label.angle=90}{bM}
		\fmfv{label=#3,label.angle=-90}{vL}
		\fmfv{label=#4,label.angle=-90}{vR}
	\end{fmfgraph*}}
}

\begin{fmffile}{see-saw}
	\fmfframe(3,3)(3,5){
		\begin{fmfgraph*}(28, 11)
			\fmfpen{0.7}
			\fmfstraight
			\fmfset{arrow_len}{2mm}
			\fmfleft{bL,tL}
			\fmfright{bR,tR}
			\fmf{phantom}{tL,miaL,tM,miaR,tR}
			\fmf{fermion,label=\niceNu{}{3},label.side=left}{bL,vL}
			\fmf{fermion}{bM,vL}
			\fmf{fermion}{bM,vR}
			\fmf{fermion,label=\niceNu{}{3},label.side=right}{bR,vR}
			\fmffreeze
			\fmf{dashes}{vL,miaL}
			\fmf{dashes}{vR,miaR}
			\fmfv{label=\niceNu{}{4},label.angle=90}{bM}
			\fmfv{label=$y$,label.angle=-90}{vL}
			\fmfv{label=$y$,label.angle=-90}{vR}
			\fmfv{d.sh=cross,label=$\left<H_1\right>$,label.angle=180,d.siz=4thick}{miaL}
			\fmfv{d.sh=cross,label=$\left<H_1\right>$,label.angle=0,d.siz=4thick}{miaR}
	\end{fmfgraph*}}
	\hfill
	\neutrinosed{\niceNu{}{2}}{\niceNu{}{2}}{$d\vphantom'$}{$d\vphantom'$}
	\hfill
	\neutrinosed{\niceNu{}{2}}{\niceNu{}{3}}{$d\vphantom'$}{$d'$}
	\hfill
	\neutrinosed{\niceNu{}{3}}{\niceNu{}{3}}{$d'$}{$d'$}
\end{fmffile}
}
\caption{
	Diagrams, contributing to the \oneLoopNuMatrix{} mass matrix. 
	Arrows show the flow of chirality. 
}
\label{fig:Neutrino-mass diagrams}
\end{figure}

%
After this diagonalization step, we take into account one-loop
corrections to the masses. Since one of the neutrinos does not couple
to the Higgs bosons, the one-loop neutrino mass corrections lead to
the block structure indicated in the third matrix of
figure~\ref{fig:neutrino-rotations}:
the two non-vanishing blocks at the level of our approximation
are the
$2\times2$ mass matrix block \oneLoopNuMatrix{}, which contains the
tree-level eigenvalue $m_3$ and  one-loop self-energy corrections, and the heavy seesaw neutrino mass $m_4$. All other entries are either generated at higher loops or are neglected in one-loop diagrams as being either proportional to $y^2 \times \text{loop}$, which is extremely small in our case (see eq.~\eqref{eq:seesaw-limit}), or to $\frac{m_3}{m_4} \times \text{loop}$, which has an additional seesaw suppression factor. 
As will be shown in the next section, the last rotation \loopRotation{2} diagonalizes \oneLoopNuMatrix{} and
can be used to parameterize the Yukawas couplings \massY{i}{} in the mass eigenstate basis.
Its form will be described in eq.~\eqref{eq:R definition}.

There are in total three non-zero masses for neutrinos at one-loop level. We write them in the ascending order as:
\begin{equation}
\hat{m} :=\text{diag}
\big(0,\, \pole{2},\, \pole{3},\, m_4\big)
,
\label{eq:pole masses}
\end{equation}
where we consider the loop corrections for the heaviest mass to be negligible.
Also, $m_{4} \gg m_{3}$ leads to
\begin{equation}
\fullRotation{4}=\loopRotation{4}\seesawRotation{4}\niceRotation{4}
\approx
\loopRotation{4}\niceRotation{4}\,
\Rightarrow \fullRotation{3}\approx \loopRotation{3}\niceRotation{3}
,
\label{eq:approximation}
\end{equation}
which means that neutrino masses and mixings can be related to the so-called
$3\nu$ mixing paradigm $|\fullRotation{4}_{i4}|\ll1$, which was indicated in the definition of \fullRotation{4} in eq.~\eqref{eq:four-four-matrices}. 
The \gls{pmns} matrix is defined as \citep{Zyla:2020zbs}:
\begin{equation}
\pmns{} =\left(\begin{matrix}1 & 0 & 0\\
0 & c_{23} & s_{23}\\
0 & -s_{23} & c_{23}
\end{matrix}\right)\left(\begin{matrix}c_{13} & 0 & s_{13}e^{-i\delta_{\text{CP}}}\\
0 & 1 & 0\\
-s_{13}e^{i\delta_{\text{CP}}} & 0 & c_{13}
\end{matrix}\right)\left(\begin{matrix}c_{12} & s_{12} & 0\\
-s_{12} & c_{12} & 0\\
0 & 0 & 1
\end{matrix}\right)\left(\begin{matrix}e^{i\eta_{1}} & 0 & 0\\
0 & e^{i\eta_{2}} & 0\\
0 & 0 & 1
\end{matrix}\right),\label{eq:pmns}
\end{equation}
where $\eta_{1}$ and $\eta_{2}$ are unknown Majorana phases, $s_{ij}=\sin\theta_{ij}$, and $c_{ij}=\cos\theta_{ij}$, see eq.~\eqref{eq:neutrino data}.
In the \gls{gnm}, the lightest neutrino has a vanishing mass and the phase $\eta_1$ can be absorbed into a redefinition of the corresponding field, as shown in eq.~\eqref{eq:phase freedom}. Hence only $\eta_{2}$ is physical.
To keep our study simple, and since there is no evidence of $\eta_{2}$ in the PMNS matrix so far, we set it to zero.
The zero-mass lightest neutrino also implies a lower bound on neutrinoless double-beta decay \cite{Reig:2018ztc}, which is however left out of the scope of this paper.

The pole masses in eq.~\eqref{eq:pole masses} are different for \gls{nh} and \gls{ih}. 
Since the mass of the lightest neutrino vanishes, the measured mass squared differences from the neutrino oscillation experiments determine the actual neutrino masses. Therefore, eq.~\eqref{eq:all rotations} connects the \gls{pmns} matrix with \fullRotation{3}, which we can summarize as:
\begin{equation}
\begin{alignedat}{3}
\text{\gls{nh}:}\quad
& m_{2}^{\text{pole}}=\sqrt{\hphantom{|}\Delta m_{21}^{2}\hphantom{|}}\,,
&& m_{3}^{\text{pole}}=\sqrt{|\Delta m_{32}^{2}|+\Delta m_{21}^{2}}\,,\
&&
\fullRotation{3}=\pmns{\dagger}
\,,
\\
\text{\gls{ih}:}\quad 
& m_{2}^{\text{pole}}=\sqrt{|\Delta m_{32}^{2}|- \Delta m_{21}^{2} }\,,\
&& m_{3}^{\text{pole}}=\sqrt{|\Delta m_{32}^{2}| }\,,
&&
\fullRotation{3}=O_{\text{\gls{ih}}}^{\vphantom{\dagger}}\pmns{\dagger}
\,,
\label{eq:to experiment}
\end{alignedat}
\end{equation}
where the sign convention for $\Delta m_{32}^{2}$ is explicitly avoided, and the mass ordering of \gls{ih} is related to the usual $3\nu$ conventions by:
\begin{equation}
O_{\text{\gls{ih}}}=\left(\begin{matrix}
0 & 1\\
\mathbb{1}_{2\times 2} & 0
\end{matrix}\right)
\,.
\end{equation}
We take as numerical values in our code from ref.~\cite{Zyla:2020zbs}:
\begin{equation}\label{eq:neutrino data}
\begin{gathered}
\Delta m_{21}^{2}=7.4\cdot 10^{-5}~\text{eV}^{2}
\,,\quad
\left|\Delta m_{32}^{2}\right|=2.5\cdot 10^{-3}~\text{eV}^{2}
\,, \\
\theta_{12}=0.59\,\text{\small rad}
\,,\quad
\theta_{23}=0.84\,\text{\small rad}
\,,\quad
\theta_{13}=0.15\,\text{\small rad}
\,,\quad
\delta_{\text{CP}}=4.5\,\text{\small rad}
\,,
\end{gathered}
\end{equation}
where $\delta_{\text{CP}}$ is taken from the intersection of measured $1\sigma$-regions for \gls{nh} and \gls{ih} scenarios.

\subsection{Neutrino mass matrix}
The Yukawa interactions introduced in eq.~\eqref{eq:lepton neutrino yukawa} can be parameterized by four real parameters, as shown in eq.~\eqref{eq:GN parameters}.
One can connect them with the definition of the unitary $2\times2$ rotation \loopRotation{2} and use the compact parameters of the latter instead.
This section provides the required relations.

The light neutrino mass matrix \oneLoopNuMatrix{} disentangles from the heavy one due to eq.~\eqref{eq:approximation}, see ref.~\citep{Grimus:2000vj} and
figure~\ref{fig:neutrino-rotations}.
The interactions with the doublet $H_{1}$ are proportional to the Yukawa coupling $y\sim |\flavorY{1}{}|$ in eq.~\eqref{eq:seesaw-limit}, which we assume to be vanishingly small due to the approximate $Z_2$ symmetry. The relevant contributions to \oneLoopNuMatrix{} are shown in figure~\ref{fig:Neutrino-mass diagrams}:
\begin{equation}
\oneLoopNuMatrix \approx
\bigg(\begin{matrix}
0 & 0 \\
0 & m_3
\end{matrix}\bigg)
+
\Lambda\,
\bigg(\begin{matrix}
d^2 & i d d^\prime \\
i d d^\prime & -d^{\prime 2}
\end{matrix}\bigg)
\,,\quad
\Lambda
:=
\frac{m_{4}}{32\pi^{2}}
\big[
B_0(0\,, m_4^2\,, m_A^2) - B_0(0\,, m_4^2\,, m_H^2)
\big]
\label{eq:mass matrix}
\end{equation}
with the loop contribution similar to the scotogenic model \cite{Ma:2006km}, and the definition of $\Lambda$ as in ref.~\cite{Avila:2021mwg}.\footnote{The expression for $\Lambda$ given in ref.~\cite{Toma:2013zsa} is
	two times larger, which looks like a direct equivalence to  eq.~(11) of ref.~\cite{Ma:2006km}. 
	However, the definitions of the Yukawa couplings in these references differ by $\sqrt{2}$, which should lead to additional factor of one half for $\Lambda$ in ref.~\cite{Toma:2013zsa}.
	Hence, there might be a typo in the neutrino Yukawa couplings in ref.~\cite{Toma:2013zsa}: they are by a factor of $\sqrt{2}$ smaller then required. 
}
The diagonalization of \oneLoopNuMatrix{} is done via the Takagi decomposition:
\begin{equation}
\loopRotation{2}^*\oneLoopNuMatrix{}\loopRotation{2}^\dagger
=
\operatorname{diag}(\pole{2}\,, \pole{3})
\,,
\label{eq:loop-diagonalization}
\end{equation}
which restricts the matrix \loopRotation{2}, as shown in appendix~\ref{app:parametrization}, due to the zero determinant of the
one-loop correction term.
For the rotation itself Murnaghan's parameterization is used:
\begin{equation}
\loopRotation{2} = \bigg(\begin{matrix}
R_{22} & -R_{32}^*e^{i\RPhaseOverall} \\
R_{32} & \hphantom{-}R_{22}^*e^{i\RPhaseOverall}
\end{matrix}\bigg)
\,,\quad\text{with}\quad
R_{22}:=\cos\Rangle\ e^{i\RPhaseInput}
\,,\quad
R_{32}:=\sin\Rangle\ e^{i\RPhaseReplaced}
\,.
\label{eq:R definition}
\end{equation}
The following ranges of angles and phases define the \emph{rotation} in a unique way:
\begin{equation}
\RPhaseOverall\,,\Rangle \in[-\pi\,, \pi)
\,,\quad
\RPhaseInput\,, \RPhaseReplaced\in\Big[-\frac{\pi}{2}\,, \frac{\pi}{2}\Big]
\,.
\label{eq:parameter ranges}
\end{equation}
The parameter ranges that uniquely describe \gls{clfv} ratios are smaller:
\begin{equation}
\Rangle\,,\RPhaseInput\in\Big(-\frac{\pi}{2}\,, \frac{\pi}{2}\Big]
\,,
\label{eq:parameter ranges for clfv}
\end{equation}
which is derived in appendix~\ref{app:parameter space}. 
\begin{figure}[t!]
	\hspace*{\fill}
	\begin{subfigure}[c]{.44\textwidth}
		\centering
		\includegraphics[height=6.7cm]{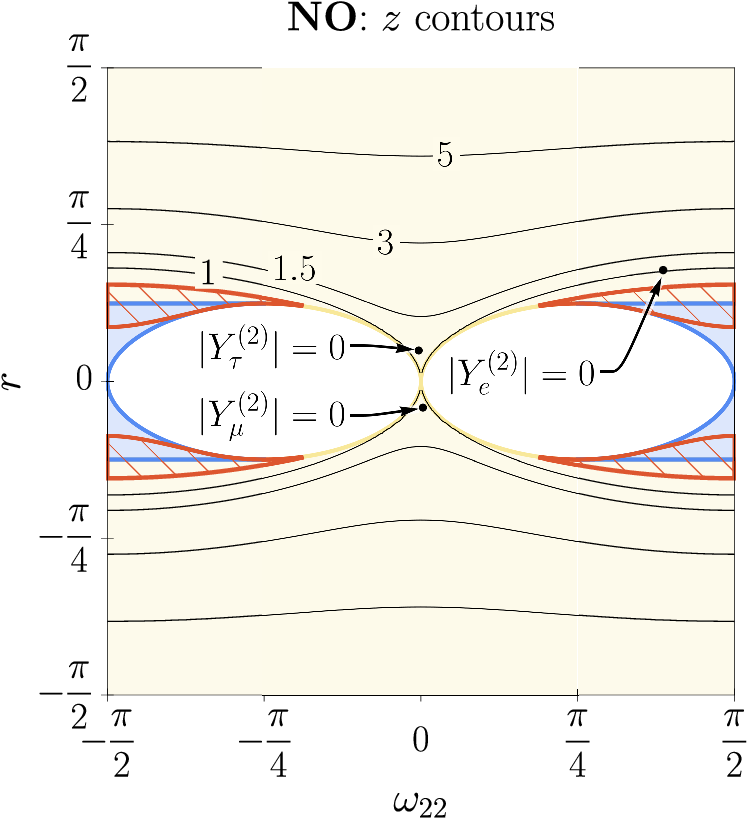}
	\end{subfigure}
	\hfill
	\begin{subfigure}[c]{.54\textwidth}
		\centering
		\includegraphics[height=6.7cm]{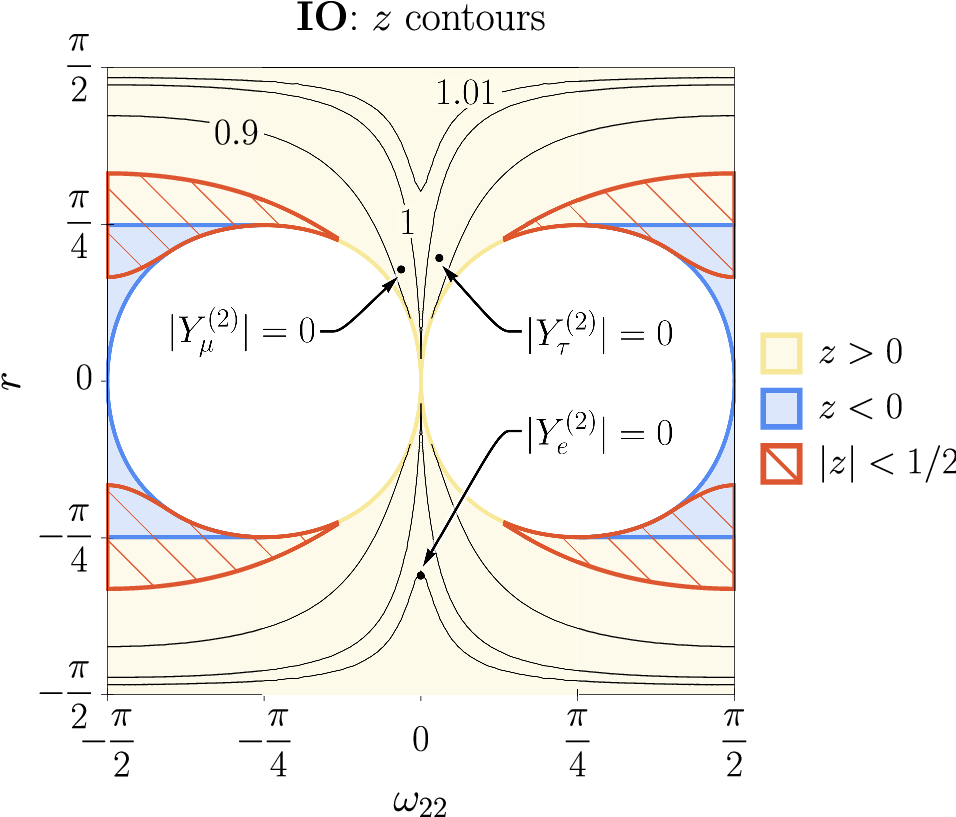}
	\end{subfigure}
	\hspace*{\fill}
	\caption{
Parameter regions of the \plane{}, representing two out of the four
physical input parameters of the model  \Rangle{}, \RPhaseInput{}, $m_4$ and $|\Lambda|$. The definition of $z$, eq.~\eqref{eq:def z}, is 
essential for the restrictions of \RPhaseInput{} and \Rangle{}, as it has
to be real, as discussed in appendix~\ref{app:parameter space}. 
The white areas are excluded by not fulfilling eq.~\eqref{eq: excluded region}. 
The contours of the values of $z$ are in the range allowed by 
eq.~\eqref{eq:consistency 2}, except for the red striped region, where
$|z|<\frac{1}{2}$, which we exclude as being fine-tuned. These red striped 
areas will also be colored white in later plots, as we will not consider 
this parameter region for the discussed exclusion criteria. 
The points where the second Yukawa coupling to electrons, muons, or taus 
vanish exactly are also shown for both hierarchies. 
	}
	\label{fig:omega-r-region}
\end{figure}

The four degrees of freedom \textemdash{} $m_3$, $d$ and complex $d^\prime$
\textemdash{} are replaced by the two neutrino pole masses \pole{2,3} and
the rotation matrix parameters \Rangle{} and \RPhaseInput{}. 
The other parameters of \loopRotation{2} are fixed by eq.~\eqref{eq:loop-diagonalization}.
In this paper we take the point of view that very large one-loop corrections for $m_3$ are
fine tuned and unnatural: hence we only analyze parameter regions
where they do not exceed 50\%, 
which corresponds to the following range for $m_3$:
\begin{equation}
m_2^\text{pole}<m_3<2 \, m^\text{pole}_3
\,.
\label{eq:consistency}
\end{equation}

It is convenient to define the abbreviation $\nuratio$, representing
the ratio between the two physical neutrino masses:
\begin{equation}
\nuratio := \frac{\pole{3}}{\pole{2}}
\quad \approx \quad
\bigg\{\begin{array}{ll}
  5.898 & \text{ for \gls{nh}} 
\\
  1.015 & \text{ for \gls{ih}}
\end{array}
\, .
\label{eq:def nuratio}
\end{equation}
It is also convenient to define the parameter $z$ by
\begin{equation}
z(\Rangle\,, \RPhaseInput\,, \RPhaseReplaced) 
= \cos^2 \Rangle\ e^{2i \RPhaseInput} +
\nuratio  \sin^2 \Rangle\ e^{2i \RPhaseReplaced}
\,.
\label{eq:def z}
\end{equation}
Using \eqref{eq:R relations} we see that $z$ corresponds to the
relative loop contribution to the neutrino mass $m_3$, and we can
express eq.~\eqref{eq:consistency}  
in terms of these parameters: 
\begin{equation}
z\in \mathbb{R}\,,\quad |z| = \frac{\pole{3}}{m_3}
\quad\Rightarrow\quad
0.5<|z|<\nuratio 
\,.
\label{eq:consistency 2}
\end{equation}

From eqs.~(\ref{eq:def z},~\ref{eq:consistency 2}) we determine \RPhaseReplaced{}.
There are three regions of possible solutions of eq.~\eqref{eq:consistency 2} for \RPhaseReplaced{} to be used in the parameterization. 
However, these solutions lead to physically equivalent Yukawa couplings, as shown in appendix~\ref{app:parameter space}, thus we use 
\begin{equation}
\RPhaseReplaced{} :=
-\frac{1}{2}\arcsin\left(\frac{\sin(2\RPhaseInput{})}{\nuratio \tan^{2}\Rangle}\right)
\,.
\label{eq:positive sol}
\end{equation}
The region for the parameters \Rangle{} and \RPhaseInput{} satisfying the constraint of eq.~\eqref{eq:consistency 2} is shown in figure~\ref{fig:omega-r-region}
as colored regions, while white regions are disallowed.
 
With these definitions, we can determine the Yukawa couplings \massY{i}{} 
in the one-loop mass eigenstate basis, which reproduce 
the \gls{pmns} matrix and the neutrino masses by construction. 
The following Yukawa couplings follow from eq.~\eqref{eq:loop-diagonalization}:
\begin{align}
\massY{2}{} &:=
\sign(\Lambda)\,
\sqrt{\frac{\pole{2}}{|z\Lambda|}}
~(0\,,  R_{22} \,, \nuratio R_{32} )
\,,
\label{eq:Y2 mass eigenstates+}
\\ 
\massY{1}{} &:=
\frac{i}{e^{i\RPhaseOverall}}
\sqrt{\frac{2\pole{3}m_4}{|z|v^2}}
~(0\,, -R_{32} \,, R_{22} )
\,.
\label{eq:Y1 mass eigenstates+}
\end{align}
Inserting eq.~\eqref{eq:positive sol} into eqs.~(\ref{eq:R definition},~\ref{eq:def z},~\ref{eq:Y2 mass eigenstates+},~\ref{eq:Y1 mass eigenstates+}) leads to a parameterization of the Yukawa couplings as functions $\massY{1}{}(\Rangle{}\,, \RPhaseInput{}\,, m_4)$ and $\massY{2}{}(\Rangle{}\,, \RPhaseInput{}\,, \Lambda)$.
For \gls{clfv}, the Yukawa coupling to the first Higgs doublet, eq.~\eqref{eq:Y1 mass eigenstates+}, can be neglected in the tiny seesaw scenario, thus only $\massY{2}{}(\Rangle{}\,, \RPhaseInput{}\,, \Lambda)$ will contribute to \gls{clfv}. 

Changing the sign of $\Lambda$ changes the phase of \massY{2}{}.
But this phase cancels in the calculation of \gls{clfv} decays. 
This means that \gls{clfv} observables restrict only the absolute value of $\Lambda$. 
The relation between mass eigenstate Yukawa couplings \massY{i}{} of eqs.~(\ref{eq:Y1 mass eigenstates+},~\ref{eq:Y2 mass eigenstates+}) and the flavor ones \flavorY{i}{} is given by:
\begin{equation}
\flavorY{i}{} = \massY{i}{}\fullRotation{3}
\,.
\label{eq:flavor to mass}
\end{equation}
The rotation \fullRotation{3} is related to the \gls{pmns} matrix in eq.~\eqref{eq:to experiment}.
With the Yukawa coupling of eq.~\eqref{eq:Y2 mass eigenstates+} this resembles the Casas-Ibarra parameterization, ref.~\cite{Casas:2001sr}, adapted for the scotogenic model, ref.~\cite{Toma:2013zsa}. 
Setting $\pole{3} \to 0$ in eq.~\eqref{eq:Y2 mass eigenstates+} realizes the exact equality.

\subsection{Vanishing \flavorY{2}{}}\label{sub:special-yukawas}

Neutrinos contribute to \gls{clfv} processes via Yukawa couplings to the second Higgs doublet, \flavorY{2}{}. 
Hence, the vanishing Yukawa coupling leads to a vanishing neutrino contribution to the process of interest.  
It turns out, there are solutions that simultaneously give a vanishing Yukawa coupling in the flavor basis and reproduce neutrino masses and mixings. 
Using eqs.~(\ref{eq:R definition},~\ref{eq:Y2 mass eigenstates+},~\ref{eq:flavor to mass}) and setting $\flavorY{2}{f}=0$ for a given flavor $f$ one gets: 
\begin{equation}
\cot( \Rangle ) e^{i (\RPhaseInput - \RPhaseReplaced) }= - \nuratio \frac{U_{3 f}}{U_{2 f}}
\,,
\label{eq:Y2 zero condition}
\end{equation}
where the solution for \RPhaseReplaced{} of eq.~\eqref{eq:positive sol} is used. 
The complex equation~\eqref{eq:Y2 zero condition} gives a single solution in \gls{nh} or \gls{ih} for each flavor $f$ (and it depends only on $\RPhaseInput$ and $\Rangle$).
These points of vanishing \flavorY{2}{} in the \plane{} are shown in figure~\ref{fig:omega-r-region} as dots with corresponding callouts, and their numerical values are given in appendix~\ref{app:Y=0 numbers}.

The solutions of eq.~\eqref{eq:Y2 zero condition} are not backed by any fundamental reasoning, are not stable and are fine-tuned. 
Nevertheless, these solutions do exist and, if the small parameter ranges around them are included, they will lead to weaker constraints on the scalar sector for \gls{clfv}.
Thus, if we discard some tiny parameter region around these points, we would make our main result, a constraint on the scalar sector from \gls{clfv}, dependent on our own choice of the size of the region. 
Instead, we choose a more agnostic approach: we include it, discuss it in more detail, and provide a suggestion of how one can define a ``more likely'' scenario away from these regions and a ``fine-tuned'' one around these regions. 
In this way, we completely cover the full parameter space of the model and do not make our main result dependent on additional assumptions. 
\subsection{Reproducing neutrino masses and mixings}
To realize the numerical scans over model parameters we used the \code{FlexibleSUSY}~\cite{Athron:2014yba, Athron:2017fvs, Athron:2021kve} spectrum-generator generator, including the extension \code{NPointFunctions}~\cite{Khasianevich:2022ess}.
It allowed us to straightforwardly implement the non-trivial mass-generation mechanism for neutrino masses in the \gls{gnm}, combining tree- and loop-level generated masses, as well as the experimentally measured \gls{pmns} mixing matrix.
In particular, we created \code{SARAH}~\cite{Staub:2015kfa, Vicente:2015zba} 
model files for the realization of the \gls{gnm} studied in this paper.
In addition, we designed the \code{FlexibleSUSY} model file incorporating the parameterization of the parameter space developed in section~\ref{sec:GN model}.\footnote{The mentioned files are available on the preprint web page for this paper.}
This working setup also resulted in an independent cross-check of the consistency of our analytical one-loop parameterization of the neutrino sector.

However, using the current implementation of the code \code{FlexibleSUSY} we spotted a bug in the neutrino pole mass calculation, where the couplings in self-energies taken from \code{SARAH} were conjugated, compared to analytical expressions both done by hand and with \code{FeynArts}~\cite{Hahn:2000kx} and \code{FormCalc}~\cite{Hahn:1998yk}. 
As a workaround, the conjugation of self-energies in the neutrino pole mass calculation in \code{FlexibleSUSY} was applied.

For the studied tiny seesaw parameter region, the elements of the \gls{pmns} matrix and the output neutrino masses at one-loop are consistent within 1\% accuracy. However, for some limiting cases that lie beyond the region of our interest, the description ceases to be accurate enough.

Since we study a tiny seesaw region, scale differences in the neutrino sector are limited only to 12 orders of magnitude ($m_4\approx 10$~GeV~vs.~$m_2 \approx 10^{-11}$~GeV). 
This helps us in numerical stability. 
In fact, we checked that for a seesaw scale of $m_4>10~\text{GeV}$, the neutrino mixing matrix in \code{FlexibleSUSY} becomes inaccurate (>1\%), while for the neutrino masses the scale is higher $m_4>100~\text{GeV}$. 
We find numerically stable and correct neutrino masses and mixings, allowing up to 1\% deviation in the output of \code{FlexibleSUSY}, for
\begin{equation}
\Lambda > \pole3 \approx 5 \cdot 10^{-11}~\text{GeV}
\,.
\label{eq:Yukawa pert}
\end{equation}
It is interesting to note that this limiting value gives the largest Yukawa coupling with value $\flavorY2{} \approx O(1)$.
Hence, the Yukawa values higher than $O(1)$ and up to a perturbativity limit do not accurately reproduce the neutrino spectrum in \code{FlexibleSUSY} for this model.
For our study, however, we will not reach this scenario since \gls{clfv} gives stronger constraints in general. 
This means that all our results are consistent, cross-checked, and reproducible with \code{FlexibleSUSY}.\footnote{After correcting the mentioned issue of Majorana pole mass calculation in \code{FlexibleSUSY}.}

\section{Charged Lepton Flavor Violating processes}\label{sec:clfv processes}
The new Yukawa interactions are not only responsible for the generation of neutrino masses, but they also give rise to \gls{clfv}.
In this section we provide the theoretical formulas for the amplitudes of two-body decays 
$l_i \to l_j\gamma$, \muec{} (valid for all nuclei), and three-body decays $l_i\to l_j l_k^{\vphantom{c}} l_k^c$.
First, we present Feynman diagrams and the amplitudes, specifying relevant and negligible ones. Later, the amplitudes are combined into the decay (and conversion) rates.
The numerical calculations are also implemented in \code{FlexibleSUSY}, using the additional extension \code{NPointFunctions}~\cite{Khasianevich:2022ess}. 

\subsection{Coefficients \label{sec:coefficients}}
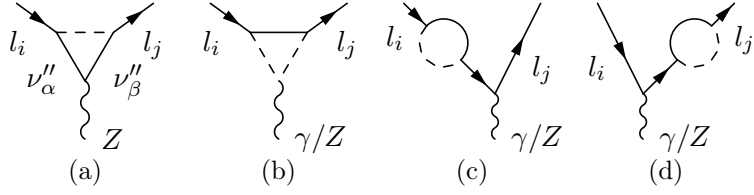
\begin{figure}[t!]
	\centering
	\begin{fmffile}{penguin-topologies}
		\unitlength = 1mm
		\begin{subfigure}[c]{.16\textwidth}
			\centering
			\begin{fmfgraph*}(18, 18)
				\fmfpen{0.7}\fmfstraight\fmfset{arrow_len}{2.5mm}
				\fmfleft{v2,v1}\fmfright{v4,v3}
				\fmf{phantom}{v2,v6,v4}
				\fmffreeze
				\fmf{fermion,label=$l_i$,label.side=right}{v1,v5}
				\fmf{fermion,label=$l_j$,label.side=right}{v7,v3}
				\fmf{boson,label=$ $,label.side=right}{v6,v8}
				\fmf{dashes,label=$ $,label.side=left,tension=0.4}{v5,v7}
				\fmf{plain,label=\massNu{}{\alpha},label.side=right,tension=0.6}{v5,v8}
				\fmf{plain,label=\massNu{}{\beta},label.side=left,tension=0.6}{v7,v8}
				\fmfv{label=$Z$, label.angle=0}{v6}
			\end{fmfgraph*}
			\subcaption{}
			\label{fig:Z-penguin1}
		\end{subfigure}
		\begin{subfigure}[c]{.16\textwidth}
			\centering
			\begin{fmfgraph*}(18, 18)
				\fmfpen{0.7}\fmfstraight\fmfset{arrow_len}{2.5mm}
				\fmfleft{v2,v1}\fmfright{v4,v3}
				\fmf{phantom}{v2,v6,v4}
				\fmffreeze
				\fmf{fermion,label=$l_i$,label.side=right}{v1,v5}
				\fmf{fermion,label=$l_j$,label.side=right}{v7,v3}
				\fmf{boson,label=$ $,label.side=left}{v6,v8}
				\fmf{plain,label=$ $,label.side=left,tension=0.4}{v5,v7}
				\fmf{dashes,label=$ $,label.side=right,tension=0.6}{v5,v8}
				\fmf{dashes,label=$ $,label.side=left,tension=0.6}{v7,v8}
				\fmfv{label=$\gamma/Z$, label.angle=0}{v6}
			\end{fmfgraph*}
			\subcaption{}
			\label{fig:Z-penguin2}
		\end{subfigure}
		\begin{subfigure}[c]{.16\textwidth}
			\centering
			\begin{fmfgraph*}(18, 18)
				\fmfpen{0.7}\fmfstraight\fmfset{arrow_len}{2.5mm}
				\fmfleft{v2,tL,pL,v1}\fmfright{v4,tR,pR,v3}
				\fmf{phantom}{v2,dummy1,v6,v4}
				\fmf{phantom}{tL,dummy2,tM,tR}
				\fmffreeze
				\fmf{boson,label=$ $,tension=5}{v6,tM}
				\fmf{fermion,label=$l_i$,label.side=right,tension=3}{v1,v5}
				\fmf{fermion,label=$l_j$,label.side=right,tension=3}{tM,v3}
				\fmf{fermion,tension=2}{v8,tM}
				\fmf{dashes,label=$ $,right,label.side=right,tension=1}{v5,v8}
				\fmf{plain,label=$ $,left,label.side=left,tension=1}{v5,v8}
				\fmfv{label=$\gamma/Z$, label.angle=0}{v6}
			\end{fmfgraph*}
			\subcaption{}
			\label{fig:Z-penguin3}
		\end{subfigure}
		\begin{subfigure}[c]{.16\textwidth}
			\centering
			\begin{fmfgraph*}(18, 18)
				\fmfpen{0.7}\fmfstraight\fmfset{arrow_len}{2.5mm}
				\fmfleft{v2,tL,pL,v1}\fmfright{v4,tR,pR,v3}
				\fmf{phantom}{v2,v6,dummy1,v4}
				\fmf{phantom}{tL,tM,dummy2,tR}
				\fmffreeze
				\fmf{boson,label=$ $,tension=5}{v6,tM}
				\fmf{fermion,label=$l_i$,label.side=right,tension=3}{v1,tM}
				\fmf{fermion,tension=2}{tM,i1}
				\fmf{fermion,label=$l_j$,label.side=right,tension=3}{i2,v3}
				\fmf{dashes,label=$ $,right,label.side=right,tension=1}{i1,i2}
				\fmf{plain,label=$ $,left,label.side=left,tension=1}{i1,i2}
				\fmfv{label=$\gamma/Z$, label.angle=0}{v6}
			\end{fmfgraph*}
			\subcaption{}
			\label{fig:Z-penguin4}
		\end{subfigure}
	\end{fmffile}
	\caption{
		Feynman diagrams for $\gamma$ and $Z$ contributions to \gls{clfv} processes for two- and three-body decays. 
		Contributions for two-body decays have an on-shell $\gamma$ on the external line, while contributions for three-body decays have off-shell $\gamma$ and $Z$ on the external line, which should be understood as subdiagrams of full penguin diagrams. 
		Arrows represent the propagation of particles; scalars and fermions lines in the loops correspond to $H^-$ and \massNu{}{4}.
		The only non-negligible contribution comes from the photon diagram in figure~\ref{fig:Z-penguin2}.
	 Figures~\ref{fig:Z-penguin3}-\ref{fig:Z-penguin4} do not give numerically meaningful contribution, but they have to be included to ensure that UV divergences cancel exactly. 
 All the diagrams that include $Z$ boson are negligible, while the diagram shown in figure~\ref{fig:Z-penguin1} is zero for a photon contribution.
	} 
	\label{fig:Z-penguin-contribution}
\end{figure}

Let us start from the simplest case of the penguin contributions, see the Feynman diagrams in figure~\ref{fig:Z-penguin-contribution}. We express the amplitude of the flavor-changing decay $l_i \to l_j\gamma$ into an
off-shell photon with \emph{outgoing} momenta $q = p_i - p_j$ as in 
ref.~\cite{Kotlarski:2019muo}:
\begin{equation}
i\Gamma_{\bar l_j l_i \gamma}
=
i \bar{u}_{j}
\Big[
\left(q^2 \gamma^\mu
-q^\mu\slashed{q}\right)\left(A_1^{L} P_L + A_1^{R}P_R
\right)
+im_{i}\sigma^{\mu\nu}q_\nu \left(
	A_2^{L}P_L+A_2^{R}P_R\right)
\Big] u_i
\,.
\label{2body-decay}
\end{equation}

For the considered scenario in the \gls{gnm}, the dominant photon contribution is given by one-loop diagrams of figures~\ref{fig:Z-penguin2}-\ref{fig:Z-penguin4} with virtual charged Higgs boson $H^-$ and virtual Majorana neutrino $\massNu{}{4}$ exchange.
In these diagrams, the lepton flavor transition is
mediated by the new Yukawa coupling $\flavorY{2}{}$.

Other contributions are negligible due to the following reasons.
The impact of Goldstone bosons $G^-_W$ is suppressed by the $Z_2$-breaking 
Yukawa coupling $|\flavorY{1}{}| \ll 1$. 
The impact of vector bosons is suppressed 
due to the \gls{gim} mechanism which leads to a $m_{\massNu{}{i}}^2/m_W^2$ factor 
and to terms proportional to $|\fullRotation{4}_{i4}| \ll 1$.
Also, in our scenario, for other \gls{clfv} processes under consideration, one can always neglect the electron Yukawa coupling \flavorY{1}{} due to its small magnitude relative to \flavorY{2}{}.

The only non-vanishing coefficients in the decay rate, eq.~\eqref{2body-decay}, are the following:
\begin{equation}
\newcommand{\myc}{\vphantom{Y}^{*\vphantom{)}}_{\vphantom{i}}}
A_1^{L} =
\frac{\flavorY{2}{i}\myc \flavorY{2}{j}}{16\pi^2 m_{H^\pm}^2}
\frac{e}{18} F_A\Big(\frac{m_4^2}{m_{H^\pm}^2}\Big)
\,,\quad
A_2^{R} =
	\frac{\flavorY{2}{i}\myc \flavorY{2}{j}}{16\pi^2 m_{H^\pm}^2 }
	\frac{e}{12}F_B\Big(\frac{m_4^2}{m_{H^\pm}^2}\Big)
\label{eq:dipole-amplitudes}
\end{equation}
with the loop functions given in eq.~\eqref{eq:loop-functions}.

Another class of \gls{clfv} diagrams are $Z$-boson penguins,  shown in
figure~\ref{fig:Z-penguin-contribution}. 
It turns out, that they can be removed from the consideration.
The diagram in figure~\ref{fig:Z-penguin1} is proportional to the
$\massNu{}{\alpha}\massNu{}{\beta} Z$ coupling. Though this coupling does not vanish for the first three generations of neutrinos, flavor-changing couplings to external leptons
include $\fullRotation{4}_{\alpha 4}$, which leads to a seesaw suppression.
The diagrams of figures~\ref{fig:Z-penguin2}-\ref{fig:Z-penguin3} have non-vanishing couplings but they lead to the zero factor of $(B_{1} + 2C_{00})$ with
\begin{equation}
B_{1} = B_{1}(0,m_4^2,m_{H^\pm}^2)
\,,\quad
C_{00} = C_{00}(0,0,0,m_{4}^{2},m_{H^\pm}^{2},m_{H^\pm}^{2})
\,.
\end{equation}
The last diagram in figure~\ref{fig:Z-penguin4} is proportional
to \flavorY{1}{}. Overall, this discussion shows that the impact of
$Z$-penguins is negligible for our scenario. Similarly, the Higgs
boson penguin is suppressed by Yukawa couplings to charged leptons. 
From this follows, that only the photon penguin gives a non-vanishing contribution.

\begin{figure}[t!]
\centering
\begin{fmffile}{box-topologies}
	\unitlength = 1mm
	\begin{subfigure}[c]{.16\textwidth}
		\centering
		\begin{fmfgraph*}(18, 18)
			\fmfpen{0.7}\fmfstraight\fmfset{arrow_len}{2.5mm}
			\fmfleft{v2,v1}\fmfright{v4,v3}
			\fmf{fermion,label=$l_i$,label.side=right,tension=3}{v1,v5}
			\fmf{fermion,label=$ $,label.side=left,tension=3}{v2,v6}
			\fmf{fermion,label=$l_j$,label.side=right,tension=3}{v7,v3}
			\fmf{fermion,label=$l_k$,label.side=left,tension=3}{v8,v4}
			\fmf{plain,label=$ $,label.side=left,tension=2}{v5,v6}
			\fmf{dashes,label=$ $,label.side=left,tension=2}{v5,v7}
			\fmf{dashes,label=$ $,label.side=right,tension=2}{v6,v8}
			\fmf{plain,label=$ $,label.side=left,tension=2}{v7,v8}
		\end{fmfgraph*}
		\subcaption{}
		\label{fig:box-mu3e1}
	\end{subfigure}
	\begin{subfigure}[c]{.16\textwidth}
		\centering
		\begin{fmfgraph*}(18, 18)
			\fmfpen{0.7}\fmfstraight\fmfset{arrow_len}{2.5mm}
			\fmfleft{v2,v1}\fmfright{v4,v3}
			\fmf{fermion,label=$l_i$,label.side=right,tension=3}{v1,v5}
			\fmf{fermion,label=$ $,label.side=left,tension=3}{v2,v6}
			\fmf{fermion,label=$l_j$,label.side=fight,tension=3}{v7,v3}
			\fmf{fermion,label=$ $,label.side=left,tension=3}{v8,v4}
			\fmf{plain,label=$ $,label.side=right,tension=2}{v5,v6}
			\fmf{dashes,label=$ $,label.side=left,tension=2}{v5,v8}
			\fmf{dashes,label=$ $,label.side=right,tension=2}{v6,v7}
			\fmf{plain,label=$ $,label.side=left,tension=2}{v7,v8}
		\end{fmfgraph*}
		\subcaption{}
	\end{subfigure}
	\begin{subfigure}[c]{.16\textwidth}
		\centering
		\begin{fmfgraph*}(18, 18)
			\fmfpen{0.7}\fmfstraight\fmfset{arrow_len}{2.5mm}
			\fmfleft{v2,v1}\fmfright{v4,v3}
			\fmf{fermion,label=$l_i$,label.side=right,tension=3}{v1,v5}
			\fmf{fermion,label=$ $,label.side=left,tension=3}{v2,v6}
			\fmf{fermion,label=$l_j$,label.side=right,tension=3}{v7,v3}
			\fmf{fermion,label=$ $,label.side=left,tension=3}{v8,v4}
			\fmf{plain,label=$ $,label.side=left,tension=2}{v5,v7}
			\fmf{dashes,label=$ $,label.side=left,tension=2}{v5,v8}
			\fmf{dashes,label=$ $,label.side=left,tension=2}{v6,v7}
			\fmf{plain,label=$ $,label.side=right,tension=2}{v6,v8}	
		\end{fmfgraph*}
		\subcaption{}
	\end{subfigure}
	\begin{subfigure}[c]{.16\textwidth}
		\centering
		\begin{fmfgraph*}(18, 18)
			\fmfpen{0.7}\fmfstraight\fmfset{arrow_len}{2.5mm}
			\fmfleft{v2,v1}\fmfright{v4,v3}
			\fmf{fermion,label=$l_i$,label.side=right,tension=3}{v1,v5}
			\fmf{fermion,label=$ $,label.side=left,tension=3}{v2,v6}
			\fmf{fermion,label=$l_j$,label.side=right,tension=3}{v7,v3}
			\fmf{fermion,label=$ $,label.side=left,tension=3}{v8,v4}
			\fmf{dashes,label=$ $,label.side=left,tension=2}{v5,v7}
			\fmf{plain,label=$ $,label.side=left,tension=2}{v5,v8}
			\fmf{plain,label=$ $,label.side=left,tension=2}{v6,v7}
			\fmf{dashes,label=$ $,label.side=right,tension=2}{v6,v8}
		\end{fmfgraph*}
		\subcaption{}
		\label{fig:box-mu3e4}
	\end{subfigure}
		\begin{subfigure}[c]{.16\textwidth}
		\centering
		\begin{fmfgraph*}(18, 18)
			\fmfpen{0.7}\fmfstraight\fmfset{arrow_len}{2.5mm}
			\fmfleft{v2,v1}\fmfright{v4,v3}
			\fmf{fermion,label=$\mu$,label.side=right,tension=3}{v1,v5}
			\fmf{fermion,label=$ $,label.side=left,tension=3}{v2,v6}
			\fmf{fermion,label=$e$,label.side=right,tension=3}{v7,v3}
			\fmf{fermion,label=$d$,label.side=left,tension=3}{v8,v4}
			\fmf{plain,label=$ $,label.side=left,tension=2}{v5,v7}
			\fmf{dashes,label=$ $,label.side=left,tension=2}{v5,v8}
			\fmf{dashes,label=$ $,label.side=left,tension=2}{v6,v7}
			\fmf{fermion,label=$u$,label.side=right,tension=2}{v6,v8}
		\end{fmfgraph*}
		\subcaption{}
		\label{fig:box-d}
	\end{subfigure}
	\begin{subfigure}[c]{.16\textwidth}
		\centering
		\begin{fmfgraph*}(18, 18)
			\fmfpen{0.7}\fmfstraight\fmfset{arrow_len}{2.5mm}
			\fmfleft{v2,v1}\fmfright{v4,v3}
			\fmf{fermion,label=$\mu$,label.side=right,tension=3}{v1,v5}
			\fmf{fermion,label=$ $,label.side=left,tension=3}{v2,v6}
			\fmf{fermion,label=$e$,label.side=right,tension=3}{v7,v3}
			\fmf{fermion,label=$u$,label.side=left,tension=3}{v8,v4}
			\fmf{dashes,label=$ $,label.side=right,tension=2}{v5,v6}
			\fmf{plain,label=$ $,label.side=left,tension=2}{v5,v7}
			\fmf{fermion,label=$d$,label.side=right,tension=2}{v6,v8}
			\fmf{dashes,label=$ $,label.side=left,tension=2}{v7,v8}
		\end{fmfgraph*}
		\subcaption{}
		\label{fig:box-u}
	\end{subfigure}
\end{fmffile}
\caption{
	Box contribution.
	The first two four-lepton diagrams from the left are proportional to $m_4^2D_0$,
	the last two of them \textemdash{} to $2D_{00}$.
	Arrows represent the propagation of particles;  scalar and fermion lines in the loops correspond to $H^-$ and \massNu{}{4}.
}
\label{fig:box-contribution}
\end{figure}
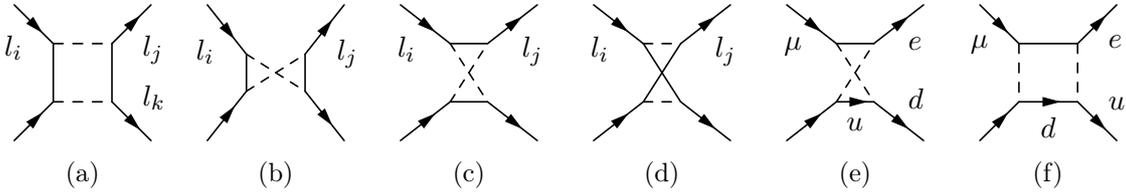

Next, we consider box contributions relevant for $l_i\to l_j l_k^{\vphantom{c}}
l_k^c$, see figures~\ref{fig:box-mu3e1}-\ref{fig:box-mu3e4}. They sum up to the result:
\begin{equation}\label{eq:A box}
\newcommand{\myc}{\vphantom{Y}^{*\vphantom{)}}_{\vphantom{i}}}
A^{LL}_{\text{box}} =
\frac{\flavorY{2}{i}\myc \flavorY{2}{j}}{16\pi^2m_{H^\pm}^2}
\frac{|\flavorY{2}{k}|^2}{2}F_C\Big(\frac{m_4^2}{m_{H^\pm}^2}\Big)
\,,\quad\text{with}\quad
\smYukawa{}{ee}\approx 0
\,,\quad\text{and}\quad
\smYukawa{}{kk} \ll \flavorY{2}{k}.
\end{equation}
 
For the \muec{} in presence of a nucleus, there are relevant box diagrams as well, see figures~\ref{fig:box-d}-\ref{fig:box-u}.
They lead to a vanishingly small contribution, because in the \gls{gnm} quarks are coupled 
to $H_1$, leading to a \flavorY{1}{} suppression in addition to the small Yukawa couplings.

The loop functions used above have the following form (see \cite{Hisano:1995cp}):
\begin{equation}
\begin{aligned}
F_A(x) &= \frac{1}{2(1-x)^4}\big(
2 - 9x + 18x^2 - 11x^3 + 6 x^3 \ln x
\big)
\,,\\
F_B(x) &= \frac{1}{(1-x)^4}\big(
1 - 6x + 3x^2 + 2x^3 - 6x^2 \ln x
\big)
\,,\\
F_C(x) &= \frac{1}{(1-x)^3} \big(1+4x-5x^2+2x(2+x)\ln x\big)
\,.
\end{aligned}
\label{eq:loop-functions}
\end{equation}
The relevant limit for us is $|x|\ll1$, and the
loop functions are normalized as $F_i(x\to0)=1$.

Using eqs.~(\ref{eq:Y2 mass eigenstates+},~\ref{eq:flavor to mass}) one obtains for $m_4 \ll v\lesssim m_{H^\pm}$ that
\begin{equation}
A_1^{L} \propto \frac{1}{\photonFactor}
\,,\quad
A_2^{R} \propto \frac{1}{\photonFactor}
\,,\quad
A^{LL}_{\text{box}} \propto \frac{1}{\boxFactor}
\,,\label{eq:propto-amplitudes}
\end{equation}
while the only other parameters that the amplitudes depend on are \Rangle{} and \RPhaseInput{}. This dependence is only slightly more complicated and can be read out from the Yukawa couplings. 
Note, that the single parameter relating the scalar sector to the \gls{clfv} two-body decays is \photonFactor{}.
This is also the most important factor for three-body decays because
contributions from box diagrams can be neglected for
$m_{H^\pm} \lesssim \text{TeV}$.
In addition, the model predicts a very simple relationship between the
two photonic amplitudes:
\begin{equation}
A_1^{L} \approx \frac{2}{3} A_2^{R}\,.
\label{eq:A1A2relation}
\end{equation}
\subsection{Decay widths}
We use the following convention for a covariant derivative:
\begin{equation}\label{eq:low-covariant-derivative}
\mathcal{D}_\mu=\partial_\mu + i e Q_f A_\mu
\,,
\end{equation}
where $e>0$ is the charge unit, $Q_f$ the electric charge of a corresponding fermion $f$ and $A_\mu$ the photon field.

The total decay width of $l_i\to l_j\gamma$ simplifies from eq.~\eqref{2body-decay} to (see, e.g. \cite{Hisano:1995cp, Kotlarski:2019muo}):
\begin{equation}
\Gamma_{l_i\to l_j\gamma} = \frac{m_i^5}{16\pi}
\norm{A_2^R}
\,.
\end{equation}
The partial decay width with three leptons of the same generation in the final state is (see \cite{Okada:1999zk,Crivellin:2017rmk,Kuno:1999jp})
\begin{equation}\label{eq:gamma-mu3e}
\begin{aligned}
\Gamma_{l_i\to 3l_j}
=
\frac{m_i^5}{192\pi^3}
\bigg[&
e^2 \norm{A_2^R} \Big(\ln\frac{m_i^2}{m_j^2} - \frac{11}{4}\Big)
+
\frac{e}{2} \operatorname{Re}\big[(2A_3^{LL}+A^{LR}_3)
A_2^{R*}\big]\\
&+
\frac{1}{4}\norm{A^{LL}_3} + \frac{1}{8}\norm{A_3^{LR}}
\bigg]
\,,
\end{aligned}
\end{equation}
where the $u$-channel for photon penguins is included via Fierz identities;
for boxes, all channels are calculated directly, and the following abbreviations are used:
\begin{equation}
A^{LR}_3 = -eA_1^L
\,,\quad
A^{LL}_3 = -eA_1^L + \frac{1}{2}A^{LL}_{\text{box}}
\,.
\end{equation}
The minus sign for the photon penguin comes from the embedding into a
four-fermion amplitude, and
the factor $1/2$ for the boxes comes from Fierz identities during the matching.

For three leptons of different generations in the final state, the expression for
the partial decay rate differs \cite{Ilakovac:1994kj}:
\begin{equation}
\begin{aligned}
\Gamma_{l_i\to l_j l_k^{\vphantom{c}} l_k^c}
=
\frac{m_i^5}{192\pi^3}
\bigg[&
e^2 \norm{A_2^R} \Big(\ln\frac{m_i^2}{m_k^2} - 3\Big)
+\frac{e}{2} \operatorname{Re}\big[(A_4^{LL}+A^{LR}_3)
A_2^{R*}\big]
\\&+
\frac{1}{8}\norm{A_4^{LL}} + \frac{1}{8}\norm{A_3^{LR}}
\bigg]
\end{aligned}
\end{equation}
with different matching for boxes leading to an absence of the additional prefactor:
\begin{equation}
A^{LL}_4 = -eA_1^L + A^{LL}_{\text{box}}
\,.
\end{equation}

For the conversion rate, we use \cite{Kitano:2002mt}
\begin{equation}
	\omega_{\mu\to e} = 4m_\mu^5\Big|\frac{1}{8}A_2^R D - eA_1^LV^{(p)}\Big|^2
\end{equation}
with dimensionless integrals $D$ and $V^{(p)}$ defined there as well.
The minus sign reflects the different definition of the photon field that comes from
the comparison of covariant derivatives.

\section{Recap of important parameters}\label{sec:strategy}
For our study we restrict the parameters:
\begin{equation}
m_4 \lesssim 10~\text{GeV}
\,,\quad
m_{H^\pm} \lesssim 1~\text{TeV}
\,.
\label{eq:parameter region}
\end{equation}
The first inequality rewrites the condition on the seesaw scale of eq.~\eqref{eq:seesaw-limit}, 
the second one leads to charged Higgs boson masses well in reach of
the LHC, and also leads to negligible box contributions in almost all of the parameter space and significantly simplifies the discussion.\footnote{See next section.}

The \gls{clfv} ratios are determined by the Yukawa couplings to the second Higgs doublet, $\flavorY2{}(\RPhaseInput{}\,,\Rangle\,, \Lambda)$, 
which also includes neutrino masses and mixings determined by the oscillation parameters. 
Since the parameter $|\Lambda|$ factors out in the amplitudes as in eq.~\eqref{eq:propto-amplitudes}, all the relevant parameters for \gls{clfv} are \RPhaseInput{}, \Rangle{}, \photonFactor{} and \boxFactor{}.
However, the last one comes from the box diagrams, which is out of the scope of this paper due to a typically negligible box contribution impact for \gls{clfv} with relatively light charged Higgs bosons.
Neglecting the box contributions, all the relevant free parameters that enter \gls{clfv} then sum up to:
\begin{equation}\label{eq:scalar-portal}
\RPhaseInput
\,,\quad
\Rangle
\,,\quad 
\photonFactor
\,.
\end{equation}
The first two are the free parameters that define the Yukawa sector:
\RPhaseInput{} is a phase that is related to CP and Majorana phases at one-loop, while
\Rangle{} parameterizes the mixing between seesaw and radiative states. 
The last parameter of eq.~\eqref{eq:scalar-portal} relates the scalar
and Yukawa sectors. We will refer to it as the \emph{photon factor}, as it is a factor in front of the amplitudes that include a photon, i.e. in front of $A_1^L$ and $A_2^R$, as shown in  eq.~\eqref{eq:propto-amplitudes}.

The photon factor, \photonFactor{}, relates the scalar sector to the Yukawa sector via \gls{clfv} and it is of most interest.
Using eq.~\eqref{eq:mass matrix}, we write it explicitly:
\begin{equation}
\photonFactor{} = \frac{m_4m^2_{H^\pm}}{32 \pi^2} \ln \left( \frac{m_H^2}{m^2_A} \right)\,,\quad m_4\ll v\,. \label{eq:photon f explicit}
\end{equation}
For the tiny seesaw region, \photonFactor{} can be generalized to the general \gls{thdm} \cite{Dudenas:2022qcw}, by inclusion of the mixing of the scalar particles in $\Lambda$, eq.~\eqref{eq:mass matrix}. 
In our study, we give bounds on the photon factor that are independent of the exact form of the scalar potential or its symmetries ($Z_2$ or $CP$ breaking or not). 
Note, however, that the $Z_2$ symmetry breaking Yukawa couplings of charged leptons to the second Higgs doublet in the Higgs basis can in principle alter the \gls{clfv} values, if included. 

It is instructive to look at the dependence on the parameters of the scalar potential for the sake of physical intuition.
The inert scalar potential of eq.~\eqref{eq:Higgs potential} in the limit of approximately degenerate  heavy scalar masses simplifies the photon factor:\footnote{An exact degeneracy would lead to $\lambda_5\to 0$ and vanishing radiative neutrino mass.}
\begin{equation}
\photonFactor{} \approx \left|\lambda_{5}\right| m_{4} \cdot\frac{v^{2}}{32\pi^{2}}\,. 
\label{eq:photon factor and lam5}
\end{equation}
The non-observation of \gls{clfv} then generally leads to a lower
bound on the photon factor \photonFactor{} (and thus $\lambda_5$) as a function of the seesaw scale. 
As was mentioned before, $\lambda_5$ is a Peccei-Quinn symmetry breaking parameter that is bounded from below by the neutrino sector alone. 
The \gls{clfv} decays then allow us to improve this bound in the tiny seesaw region. 

As a last note, we stress again that the \gls{clfv} rates are related to the scalar sector only via a single parameter, \photonFactor{}. 
Hence, our results do not depend on how exactly all the parameters in the potential are realized. 
While we used the scalar potential of the \gls{idm}, motivated by an approximate $Z_2$ symmetry, and assumed that the values of scalar masses are close to each other to get the specific interpretation of a \photonFactor{} in eq.~\eqref{eq:photon factor and lam5}, 
one is, in general, free to use any scenario of the \gls{thdm} potential, as already mentioned before. 
To construct a consistent potential, one can look at e.g. \cite{Branco:2011iw,Arbey:2017gmh, Boto:2020wyf, Davidson:2005cw, Haber:2006ue, Haber:2010bw} for detailed studies of \gls{thdm} or at \cite{Kalinowski:2018ylg} for constraints in \gls{idm} specifically.

\section{Phenomenological analysis}\label{sec:pheno}
\subsection{Discussion of the relative importance of different decay modes}
\begin{table}[t!]
	\centering
	\begin{tabular}{| c | c | c |}
		\hline
		Observable & Experiments and constraints \\
		\hline
		\mueg &
		MEG~\cite{MEG:2020zxk}: $4.2\cdot 10^{-13}$
		$\rightarrow$ 
		MEG-II~\cite{MEGII:2018kmf}: $6\cdot 10^{-14}$
		\\
		\taueg & 
			BaBar~\cite{BaBar:2009hkt}: $3.3\cdot10^{-8}$
			$\rightarrow$
			Belle-II~\cite{Belle-II:2018jsg}: $3.0\cdot10^{-9}$
		\\
		\taumug &
			BaBar~\cite{BaBar:2009hkt}:
			$4.5\cdot10^{-8}$
			$\rightarrow$
			Belle-II~\cite{Belle-II:2018jsg}: $1.0\cdot10^{-9}$
		\\
		\mueee
		& 
		SINDRUM~\cite{SINDRUM:1987nra} : $1\cdot 10^{-12}$
		$\rightarrow$
		Mu3e-I~\cite{Wasili:2020ksf} : $2\cdot 10^{-15}$
		\\
		\taueee &
			Belle-I~\cite{Hayasaka:2010np}: $2.7\cdot10^{-8}$
		$\rightarrow$
			Belle-II~\cite{Belle-II:2018jsg}: $4.6\cdot10^{-10}$
		\\
		\taumee &
			Belle-I~\cite{Hayasaka:2010np}: $1.8\cdot10^{-8}$
			$\rightarrow$
			Belle-II~\cite{Belle-II:2018jsg}: $3.1\cdot10^{-10}$
		\\
		\tauemm&
			Belle-I~\cite{Hayasaka:2010np}: $2.7\cdot10^{-8}$
			$\rightarrow$
			Belle-II~\cite{Belle-II:2018jsg}: $4.6\cdot10^{-10}$ 
		\\
		\taummm& 
			Belle-I~\cite{Hayasaka:2010np}: $2.1\cdot10^{-8}$
			$\rightarrow$
			Belle-II~\cite{Belle-II:2018jsg}: $3.6\cdot10^{-10}$ 
		\\
		\muec &
		\textemdash{}
		$\rightarrow$
		COMET~\cite{COMET:2018auw}: $7\cdot 10^{-15}$
		\\
		\hline
	\end{tabular}
	\caption{Current and planned experimental bounds, related to corresponding observables. Data for $\tau$ decays for Belle-II was obtained from the figure~189 of ref.~\cite{Belle-II:2018jsg}. For the \muec{}, we consider the Al nucleus.} 
	\label{tab:observables-experiments}	
\end{table}

All the \gls{clfv} experiments, which potentially can constrain the \gls{gnm}, are shown in table~\ref{tab:observables-experiments}. 
Below, we discuss the importance of all these decay modes to single out the most constraining experiments for the \gls{gnm}.

Typically, the branching ratios for three-body decays 
are dominated by photonic contributions, while others \textemdash{} boxes, $Z$ and Higgs penguins \textemdash{} are negligible
in the parameter region defined by eq.~\eqref{eq:parameter region}. 
The tiny seesaw scale, defined as $m_4 \ll v\lesssim m_{H^\pm}$, further leads
to a fixed ratio of $A_1^{L} \approx 2/3 A_2^{R}$.
We call this regime \emph{photon} dominance (in contrast to dipole dominance, which assumes  $A_1^{L} \ll A_2^{R}$ ):
\begin{equation}
\br(l_i \to 3 l_j) \approx  \bigg[ -\frac{5\cdot \alpha}{18 \pi} +\frac{\alpha }{3 \pi } \bigg( -\frac{11}{4} + \ln\frac{m^2_i}{m^2_j} \bigg)
\bigg]\cdot \br(l_i \to l_j \gamma)
\,,
\label{eq:photon dominance alt}
\end{equation}
where $\alpha=e^2/(4 \pi)$. 
The first term of eq.~\eqref{eq:photon dominance alt} is a correction to the dipole dominance. This correction amounts to $O(10\%)$ for \mueee{}. 

The box contribution can be increased for lower $|\Lambda|$ values, as follows from eq.~\eqref{eq:propto-amplitudes}. 
However, two-body decays constrain the \photonFactor{} factor from below. 
Taking this minimum value of \photonFactor{}, a lower value for $|\Lambda|$ translates into a higher value for  $m_{H^\pm}$.
We concentrate on the lighter masses of the charged Higgs in the paper, see eq.~\eqref{eq:parameter region}, where deviations from eq.~\eqref{eq:photon dominance alt} are small.

We find that the box contributions are generally negligible in all of the \plane{} for \mueee{}, except for the very small region, where the rate for 
\mueg{} drops sharply close to the point where $\flavorY2{\mu}=0$. 
This is because the constraint on \photonFactor{} from \mueg{} is much looser around that parameter point than in all the rest of the parameter region. 
Since it is contained in a tiny enough area in the \plane{} and both, \br(\mueee{}) and \br(\mueg{}) go to zero at $\flavorY2{\mu}=0$, it gives negligible modifications to the information that the two-body decays provide.  

The only parameter region where \taummm{} can be expected to be observed
in Belle-II is around  $\flavorY2{e}=0$; such an observation can happen only if
\taumug{} is also seen.
The value of \taummm{} can in fact deviate more significantly from photon dominance, but we find that the current and planned experimental sensitivities still leave this process phenomenologically irrelevant. 
Other three-body processes are not expected to be seen in Belle-II at all.  
As a result of these checks, we conclude that in the parameter region of our study, the three-body decays are of minor importance compared to the two-body ones.

The second phases of experiments, Mu3e-II and COMET-II, will enhance the importance of the corresponding processes. 
Their branching ratios will be fixed by the photon dominance contributions in the studied parameter region as box contributions can be neglected. 
In the present paper, we do not consider these longer-term improvements. For the
next achievable improvement of
\muec{} at COMET~\cite{COMET:2018auw} the following relation holds:
\begin{equation}
	\frac{\rate(\mu\to e)}{\br(\mueg)} < \frac{\rate(\text{COMET})}{\br(\text{MEG})}
	\,,
\end{equation}
which makes it less restrictive than \mueg{}.

From now on, we will concentrate on the study of two-body decays as the most constraining decay modes for the \gls{gnm}.
Note, however, that two-body decays can vanish if the corresponding Yukawa couplings vanish.
These points in the parameter space do exist, as described in section~\ref{sub:special-yukawas}.  
Nevertheless, as can be seen from figure~\ref{fig:omega-r-region}, there is no such point in the \plane{} in which two of the Yukawa couplings in flavor basis vanish simultaneously.
This means that all three two-body decay experiments have to be combined to give a strict bound on \photonFactor{} and, hence, we further study all three of them.
\subsection{Current and planned restrictions on \photonFactor{}}
\begin{figure}[t!]
	\hspace*{\fill}
	\begin{subfigure}[c]{.49\textwidth}
		\centering
		\includegraphics[width=\textwidth]{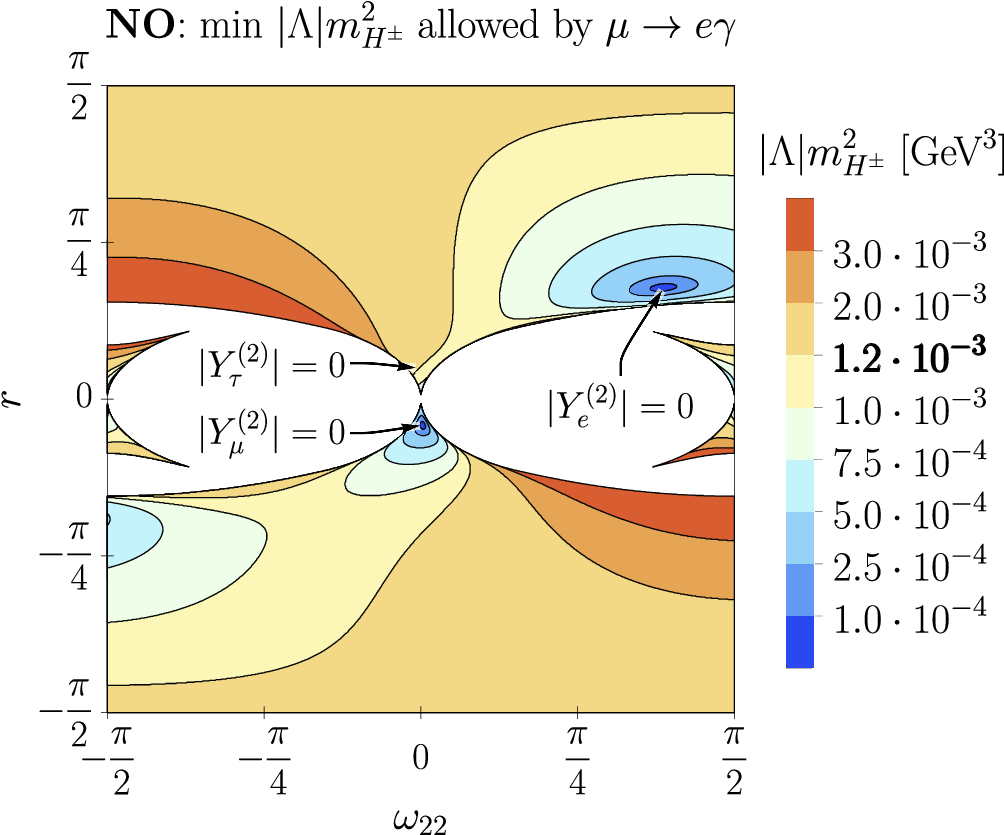}
		\caption{MEG contours}
		\label{fig:no-meg-contours}
	\end{subfigure}
	\hfill
	\begin{subfigure}[c]{.49\textwidth}
		\includegraphics[width=\textwidth]{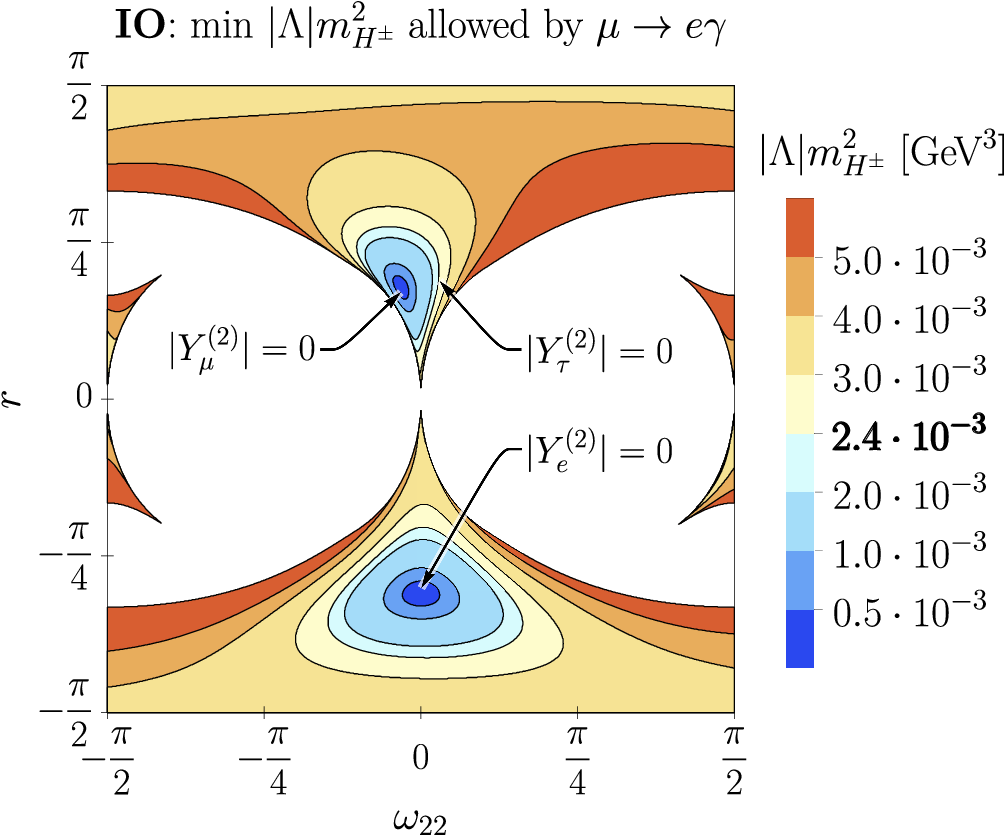}
		\caption{MEG contours}
		\label{fig:io-meg-contours}
	\end{subfigure}
	\hspace*{\fill}
	\caption{
		Contour plots of lower limits of \photonFactor{} for the current bounds on  \mueg{} by the MEG experiment. The white area is excluded theoretically, either by the constraint that $z$, eq.~\eqref{eq:def z}, has to be real, or by the constraint in eq.~\eqref{eq:consistency 2}. The latter constraint is colored red striped in figure~\ref{fig:omega-r-region}. 
		The tiny regions around the points with $\flavorY{2}{e}=0$ and $\flavorY{2}{\mu}=0$ are always allowed by \mueg{}. They are constrained either by \taumug{} or \taueg{}, as in table~\ref{tab:special point limits}.
		The bold values of the photon factor correspond to the
                critical cases when \mueg{} allows the point with $\flavorY{2}{\tau}=0$.
	}
	\label{fig:current constraints}
\end{figure}
\begin{table}[t!]
\centering
\begin{tabular}{c|c|c}
Process and parameter point &  \gls{nh}, \photonFactor{} [GeV$^3$] & \gls{ih}, \photonFactor{} [GeV$^3$]\\
\hline
\taueg{} at $\flavorY2{\mu} = 0$ & $1.9 \cdot 10^{-6}$ &  $4.0 \cdot 10^{-6}$ \\
\hline
\taumug{} at $\flavorY2{e} = 0$ & $1.3 \cdot 10^{-5}$ & $7.6 \cdot 10^{-6}$ \\
\end{tabular}
\caption{Lower bounds on the photon factor, \photonFactor{}, from $\tau$ decays at the parameter points, where $\br(\mueg{})$ vanishes. 
\label{tab:special point limits} }
\end{table}

The lower bound for \photonFactor{} from non-observation of \gls{clfv} depends on the parameters \Rangle{} and \RPhaseInput{}.
The current lower bounds from \mueg{} on \photonFactor{} for \gls{nh} and \gls{ih} are given as contour plots in figure~\ref{fig:current constraints}.
The most important observable is \mueg{}. It gives the tightest bounds almost everywhere, except for the sharp minima around the points, where the corresponding Yukawa couplings vanish.
In those small regions, either \taueg{} or \taumug{} constrains the photon factor \photonFactor{}, which results, for the current experimental limits (first column of table~\ref{tab:observables-experiments}), in the lower bounds on the photon factor, shown in table~\ref{tab:special point limits} 
(also see figure~\ref{fig:all-pessimist-3} for current and planned future bounds). 

If no \gls{clfv} processes are observed in the planned experiments listed in the second column of table~\ref{tab:observables-experiments}, one obtains improved bounds on the photon factor \photonFactor{} by the following scaling:

\begin{equation}
\big[\photonFactor{}\big]_{planned} = \sqrt{\frac{\br{}_i^{current}} {\br{}_i^{planned}}}
\big[\photonFactor{}\big]_{current} 
\,.
\label{eq:factor-bounds-scaling}
\end{equation}

\subsection{Some processes are observed: restrictions on the \plane{}} \label{sec:optimistic}
\begin{figure}[t]
	\begin{subfigure}[t]{.4\textwidth}
		\centering
		\includegraphics[height=5cm]{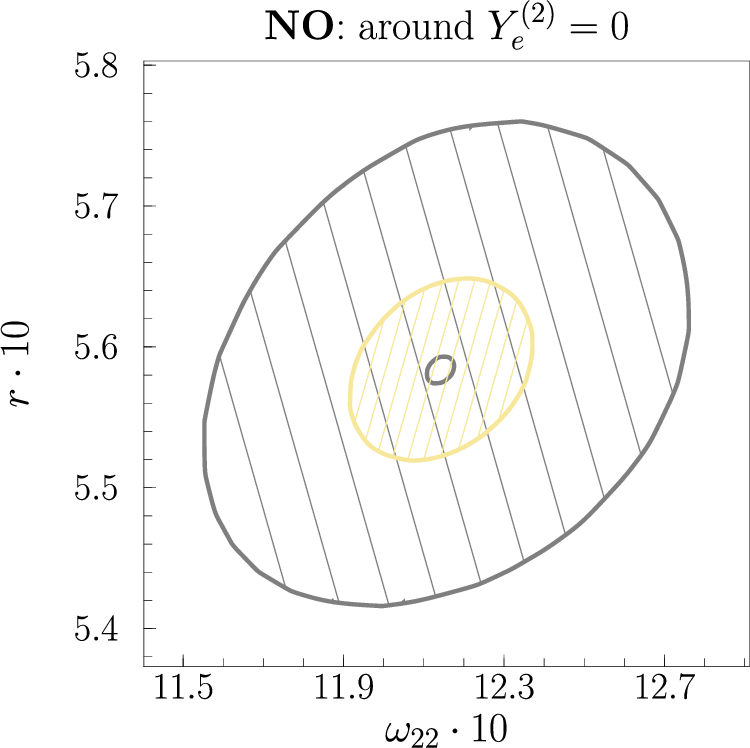}
	\end{subfigure}
	\hfill
	\begin{subfigure}[t]{.59\textwidth}
		\centering
		\includegraphics[height=5cm]{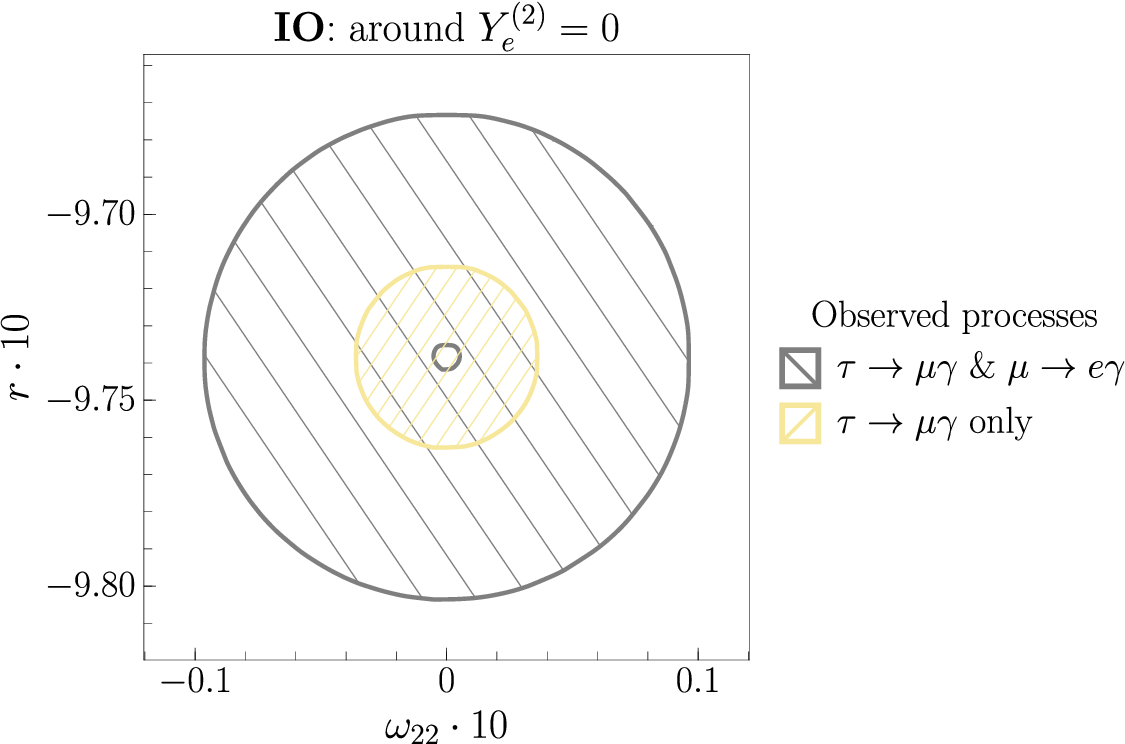}
	\end{subfigure}

	\vspace{0.5cm}
	\begin{subfigure}[t]{.4\textwidth}
		\centering
		\includegraphics[height=5cm]{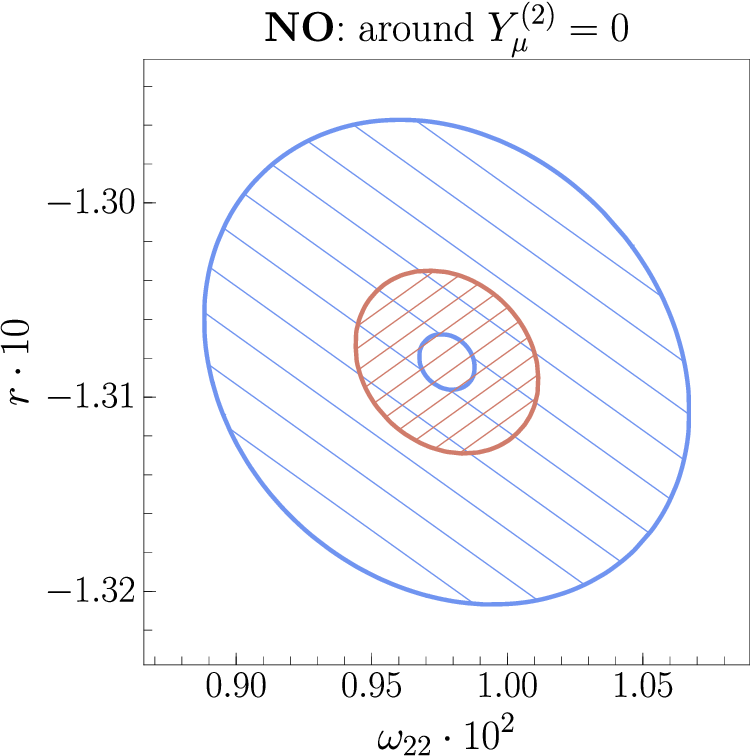}
	\end{subfigure}
	\hfill
	\begin{subfigure}[t]{.59\textwidth}
		\centering
		\includegraphics[height=5cm]{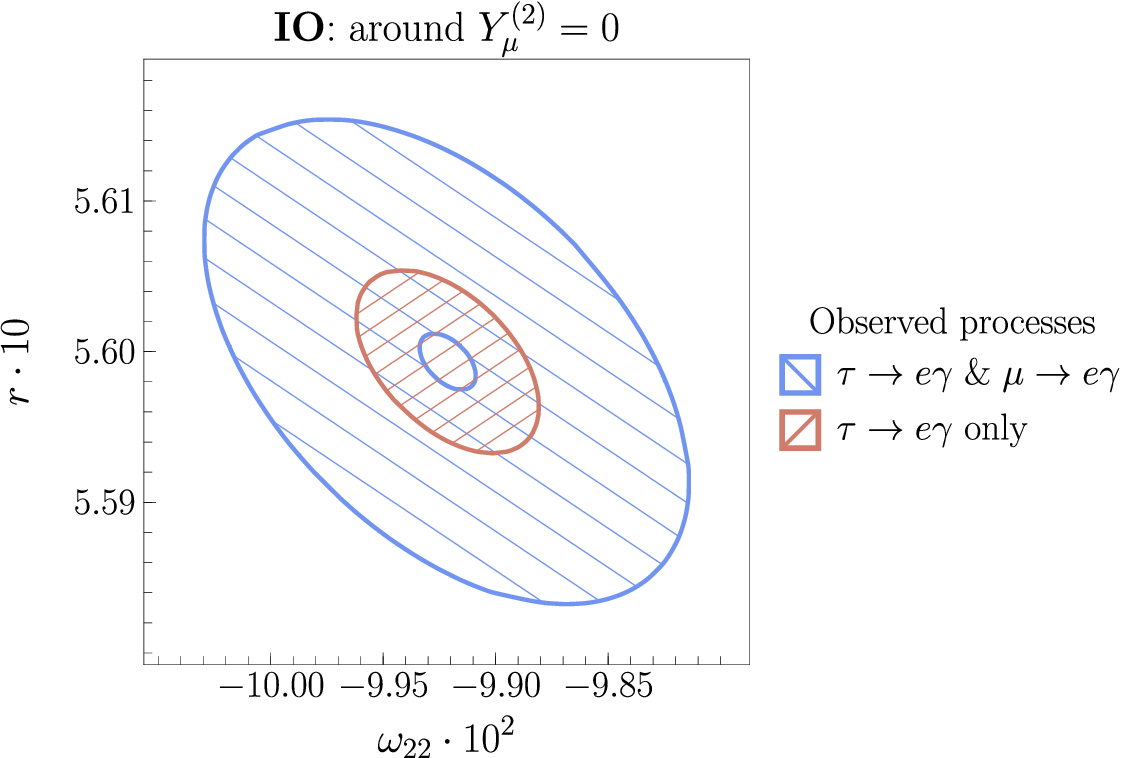}
	\end{subfigure}
	\caption{
		Regions in the \plane{} where the observation of the listed processes is possible in the planned experiments. 
		For example, the gray dashed regions in the upper plots correspond to the case, where \taumug{} and \mueg{} are observed simultaneously, the yellow dashed region shows the allowed values when only \taumug{} is observed, but \mueg{} not, as noted in the plot legend.
		Note, that the top-left area is significantly larger than other areas.}
	\label{fig:optimistic regions}
\end{figure}
\begin{table}[t]
\centering
\newcommand\mr[1]{\multirow{2}{*}{$#1$}}
\begin{tabular}{l|c|c}
Observed processes   & \gls{nh}, \photonFactor{} [GeV$^3$] & \gls{ih}, \photonFactor{} [GeV$^3$] \\
\hline
\taumug{} \& \mueg{} & $9.43\cdot 10^{-5}$ 		 & \mr{5.12\cdot 10^{-5}} 	  \\
\taumug{}~only 		 & $9.07\cdot 10^{-5}$ 		 &  						  \\
\hline
\taueg{} \& \mueg{}  & \mr{6.28\cdot 10^{-6}}    & \mr{1.34\cdot 10^{-5}}	  \\
\taueg{}~only 		 & 							 & 
\end{tabular}
\caption{Upper bounds on the photon factor, \photonFactor{}, if the only specified combination of processes is observed in Belle-II and MEG-II. Each bound corresponds to a region in the \plane{} in figure~\ref{fig:optimistic regions}.
\label{tab:optimistic limits from above}}
\end{table}

If the planned experiments with improved sensitivity do observe a signal, then the \gls{gnm} becomes very restrictive.
We use table~\ref{tab:observables-experiments} to define regions with respect to the possible outcomes of the experiments:
\begin{equation}
\br{}_i^{planned} < \br{}_i^{observed} < \br{}_i^{current}
\,,
\end{equation}
where superscripts ``current'' and ``planned'' are again used for the reach-in branching ratio of the current or planned experiments. The corresponding values are given in the first and second columns of table~\ref{tab:observables-experiments}.

The parameter regions where the observation of \taueg{} and \taumug{} in the \plane{} is possible are shown in figure~\ref{fig:optimistic regions}.
They also provide the maximal allowed photon factor by the values given in table~\ref{tab:optimistic limits from above}.
The lower bounds from table~\ref{tab:special point limits} together with the upper ones from table~\ref{tab:optimistic limits from above} specify the allowed range for the photon factor \photonFactor{} in case of these observations, while figure~\ref{fig:optimistic regions} shows the allowed regions in the \plane{}: the dashed areas. 
These dashed areas are disjoint, meaning that the \gls{gnm} predicts, that \taueg{} and \taumug{} cannot be seen together.
If \mueg{} is observed, then figure~\ref{fig:current constraints}, scaled with eq.~(\ref{eq:factor-bounds-scaling}) for the observed branching ratio, gives the values of \photonFactor{} in the \plane{}.
\subsection{Nothing is observed: future absolute bounds on \photonFactor}\label{sec:pessimist}
\begin{figure}[t!]
	\hspace*{\fill}
	\begin{subfigure}[c]{.48\textwidth}
		\centering
		\includegraphics[width=\textwidth]{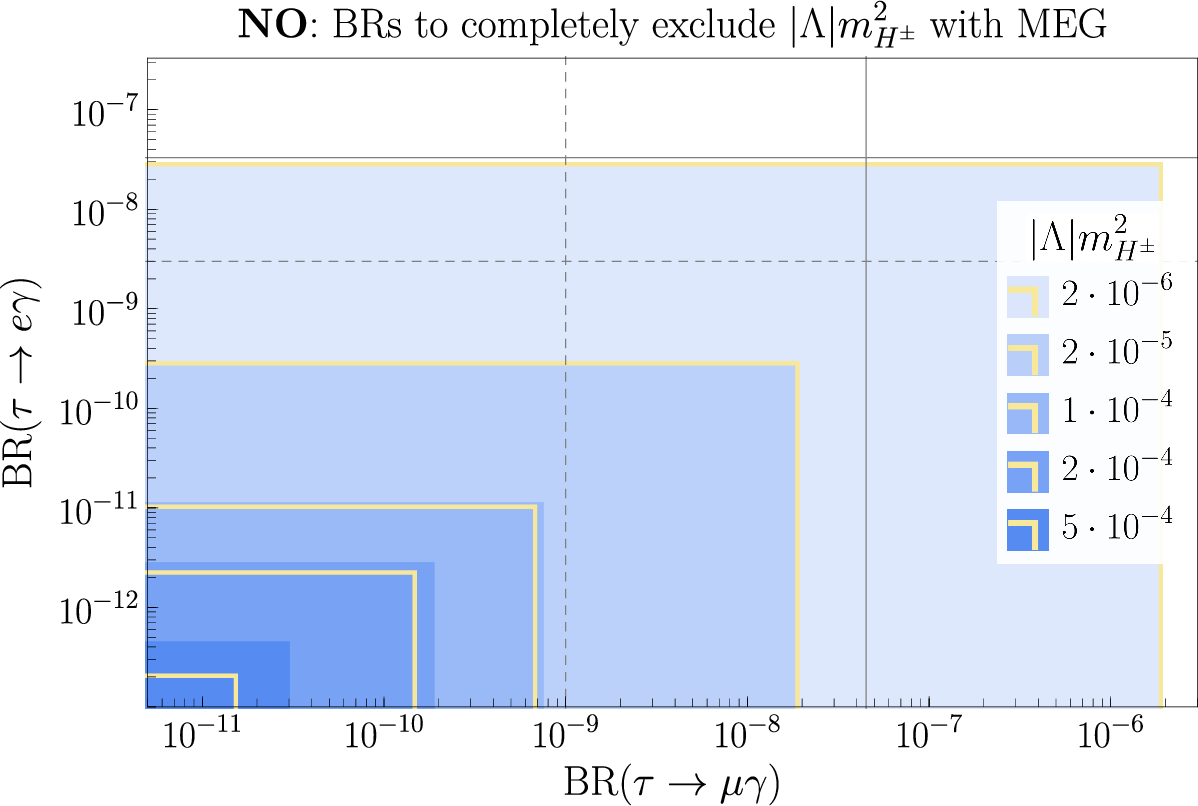}
	\end{subfigure}
	\hfill
	\begin{subfigure}[c]{.48\textwidth}
		\includegraphics[width=\textwidth]{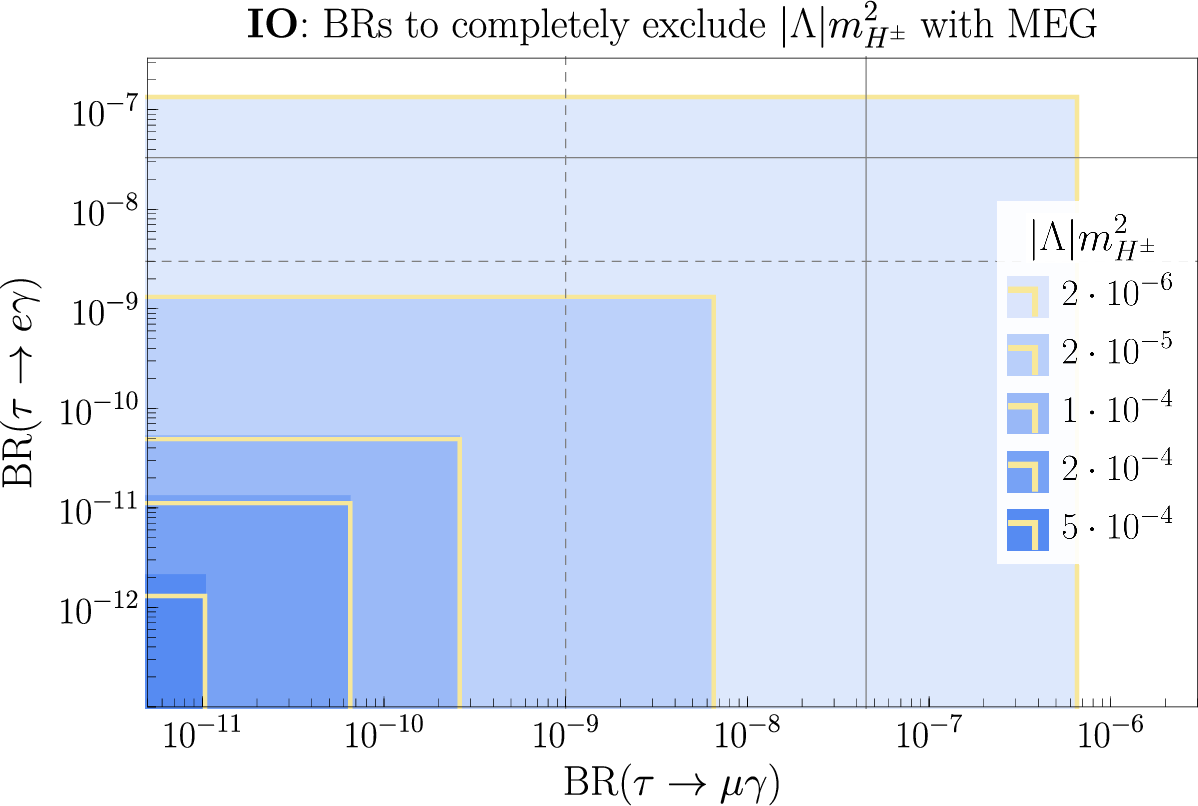}
	\end{subfigure}
	\hspace*{\fill}
	\caption{
		Branching ratios for $\tau$ decays within yellow rectangular border allow to exclude the whole \plane{} for a specified photon factor \photonFactor{} with MEG bounds on \mueg{}.
		Closest blue rectangles correspond to the case of infinitely precise experimental bound.
		To exclude larger photon factor values, bounds on \mueg{} have to be improved leading to the same rectangles up to an appropriate scaling of \photonFactor{} and axes, as mentioned in the text.
		Gray solid lines correspond to current experimental bounds, gray dashed lines \textemdash{} to the planned ones.
	}
	\label{fig:all-pessimist-3}
\end{figure}

To generalize our results and to give a more definite answer of how the combination of the neutrino sector and \gls{clfv} constrains the scalar sector, we look at the absolute bound on the photon factor, \photonFactor{}.
By absolute bound we mean the minimum value of \photonFactor{} such that the whole \plane{}, consistent with neutrino masses and mixings, is excluded by \gls{clfv} observables.
The absolute limits for current experimental bounds are shown in the first row of table~\ref{tab:special point limits} and are determined by \taueg{} experiments at $\flavorY2{\mu}=0$.
The limits for the planned experiments are derived with the scaling of eq.~\eqref{eq:factor-bounds-scaling}.

In general, \mueg{} is the strongest experiment in most of the \plane{} in the nearest future.
However, the allowed minimal \photonFactor{} is mainly determined by the weaker bounds on \taueg{} or \taumug{}.
This is because one always has two points in the parameter space where $\flavorY2{e,\mu}=0$ and, hence, the branching ratio for \mueg{} vanishes. 
In order to further improve the absolute bound on \photonFactor{} one
therefore has to improve the experimental sensitivities on \taueg{}
and \taumug{}.
The combination of branching ratios of these experiments, required to
completely exclude the \plane{} for specific values of \photonFactor{}, is shown in figure~\ref{fig:all-pessimist-3}.

The yellow borders specify precise values of branching ratios that
have to be applied, i.e.\ if the future  \taueg{}
and \taumug{} sensitivities are inside one particular yellow
rectangle, the absolute lower bound on \photonFactor{} is at least as
good as indicated in the legend.
The yellow borders are obtained without approximations and are determined by varying the branching ratio of the $\tau$ decays over the regions in the \plane{} that are not excluded by \mueg{}.
The blue rectangles are similar, but they show the hypothetical case in which \mueg{} is of infinite precision, and hence only the points with $\br(\mueg{})=0$ are allowed by \mueg{}.
All future improvements in \mueg{} experiments for a specific value of \photonFactor{} reside in the area between the closest yellow and blue rectangles.
One can see, that the size of the area between the rectangles is small
(hence for this analysis improvements in \mueg{} are of minor
importance) but grows with the values of the photon factor.

This behavior continues to higher values of \photonFactor{} than shown
in the plot, up to the moment when \mueg{} allows the point where $\flavorY{2}{\tau}=0$, as shown as critical case in figure~\ref{fig:current constraints}.
To exclude even larger \photonFactor{} values,  the improvement of bounds for \mueg{} is required.
For a different \mueg{} precision, figure~\ref{fig:all-pessimist-3} can be used with the following rescaling of the photon factor and axes:
\begin{equation}
	\photonFactor{} \rightarrow \photonFactor{} \sqrt{\frac{\br(\text{MEG})}{\br(\text{\mueg})}}
	\,,\quad
	\br(\tau) \rightarrow \br(\tau) \frac{\br(\text{\mueg})}{\br(\text{MEG})}
	\,.
\end{equation}

However, a high precision of $\tau$ experiments is required so that this case lies out of the bounds of figure~\ref{fig:all-pessimist-3}.
Taking into account the current and planned bounds for \taueg{} and \taumug{}, one concludes that only \taueg{} improvements are important for the absolute bound in the near future.

\subsection{Interpretation of results: example of limits on $\lambda_5$ and relations to scoto-seesaw and scotogenic models}\label{sec:lam5}
The absolute bounds on the photon factor, \photonFactor{}, correspond to the values of the photon factor, for which the entire \plane{} is excluded.
They are given by \taueg{} in table~\ref{tab:special point limits} or figure~\ref{fig:all-pessimist-3}.  
However, figure~\ref{fig:current constraints} shows that the absolute bounds are due to very small areas in the parameter space.
These areas are characterized by the vanishing of one of the Yukawa couplings, and only in those regions an observation of $\tau$ decays in Belle-II can be expected.
We define the ``essential part'' of the \plane{}, as the region where the observation of $\tau$ decays in Belle-II is not possible i.e. all the plane except the regions shown in colors in figure~\ref{fig:optimistic regions}.
We then obtain a ``typical'' bound on the photon factor, i.e. the bound for which the ``essential part'' of the \plane{} is excluded.
The absolute bound is shown in table~\ref{tab:optimistic limits from above} and together with this typical bound, defined above, reads as:
\begin{equation}
\begin{aligned} 
&\text{absolute for \gls{nh}(\gls{ih}):}& &\photonFactor> 1.9 (4.0) \cdot 10^{-6}\  \text{GeV}^3\,, \\ 
&\text{typical (no }\tau \to e (\text{or }\mu) \gamma \text{ expected):}& &\photonFactor \gtrsim 10^{-4}\ \text{GeV}^3 \,. \label{eq:main result}
\end{aligned} 
\end{equation}
To get an intuition of how strong this current bound on the scalar sector actually is, we can go back to the special case of the scalar sector, where eq.~\eqref{eq:photon factor and lam5} holds.
Then eq.~\eqref{eq:main result} translates into bounds on $|\lambda_5|$: 
\begin{equation}
\begin{aligned} 
&\text{absolute for \gls{nh}(\gls{ih}):}&&|\lambda_5| > 1(2)\cdot 10^{-2} \frac{\text{keV}}{m_4}\,,  \\ 
&\text{typical (no }\tau \to e (\text{or }\mu) \gamma \text{ expected):}&& |\lambda_5| \gtrsim \frac{\text{keV}}{m_4}\,. \label{eq:main result l5}
\end{aligned}
\end{equation}

As we mentioned in the introduction, the \gls{gnm} in the tiny seesaw scale region is similar to the scoto-seesaw and scotogenic models due to an approximate $Z_2$ symmetry. 
To put our results in a more general framework let us consider how our results can be applied to these two models. 

The scoto-seesaw model has two heavy neutrinos, one of which is odd under the $Z_2$ symmetry and the other one is even.
One can use our parameterization for the Yukawa couplings, where $m_4$ in \flavorY2{} is understood as the mass of the $Z_2$-odd neutrino, while $m_4$ in the expression for \flavorY1{} has to be understood as the mass of the $Z_2$-even neutrino of the scoto-seesaw model. 
In case these heavy neutrinos of the scoto-seesaw model are lighter than the scalars, only \flavorY2{} enters the \gls{clfv} just as in our case, and hence all our results for \gls{clfv}, including all figures, hold for the scoto-seesaw model, too.  
We also note that our parameterization allows to cover the full parameter region of the scoto-seesaw model, while \cite{Rojas:2018wym, Mandal:2021yph} assume only scenarios, where the mixing between the radiative and the seesaw neutrino at one-loop is absent, i.e. $\Rangle=0$ in our parameterization.

The scotogenic model also has a $Z_2$ symmetry, yet it does have more parameters. 
Nevertheless, the Casas-Ibarra parameterization gives the same general behavior of the Yukawa couplings with respect to \photonFactor{} and hence \gls{clfv} ratios like in our case. 
The difference is that in the scotogenic model \photonFactor{} is a $3\times 3$ matrix and, instead of two parameters, \Rangle{} and \RPhaseInput{}, one has a general $3\times 3$ orthogonal matrix entering the Yukawa couplings. 
Nevertheless, the same \emph{typical} behavior for the tiny seesaw scales in the scotogenic model is to be expected. 
One should also have in mind that in the scotogenic model there could be similar cancellations suppressing the \mueg{} branching ratio like in our case. These cancellations would push the absolute bound on \photonFactor{} lower.
However, from our study we cannot generalize our strict exclusions and predict them for the scotogenic model. 

Even though the tiny seesaw scale has never been rigorously studied in the scotogenic model before, we do find the comment about it in \cite{Toma:2013zsa}, which gives us some way to put our results in a more general context.
\cite{Toma:2013zsa} claims that for $m_4 \ll m_{H^\pm}$ the value $\lambda_5=10^{-9}$ is excluded. 
We see that the typical bound, given in eq.~\eqref{eq:main result l5}, indeed is consistent with this statement for scalar masses of the TeV scale, which are the masses they actually consider. 
This confirms our expectation that the typical behavior of the models in this parameter region is the same.  
As said before, our absolute bound from eq.~\eqref{eq:main result l5} cannot be directly applied to the scotogenic model, yet it gives an example of how much special analytical solutions, that can hardly be caught in random parameter scans, can push the strict exclusion limits of the model.

\section{Conclusions}\label{sec:conclusions}
Throughout this paper, we concentrated on the question: how does the
neutrino sector of the \gls{gnm}  constrain the scalar sector
via \gls{clfv} observables? The \gls{gnm} is a model where neutrino
masses are generated with a minimal neutrino sector but a non-minimal
Higgs sector, such that new Yukawa couplings can lead to \gls{clfv} effects.
Guided by the principle ``exclude as much as possible'', we singled out the scenario where the strongest constraints can be drawn, i.e. when \gls{clfv} decays are enhanced. 
This scenario corresponds to a tiny seesaw scale (lower than the
electroweak scale) and a mass of the charged Higgs $m_{H^\pm}$ slightly above the
electroweak scale (up to a TeV). 

In this parameter range, box diagrams are negligible, and the study
simplifies to the relations between three parameters: two
parameters, \RPhaseInput{} and \Rangle{}, parameterize the new Yukawa
sector, see figure~\ref{fig:omega-r-region}, and one parameter, the so-called photon factor \photonFactor{}, includes scalar sector parameters and the seesaw scale, see eq.~\eqref{eq:photon f explicit}. 

The non-observation of \gls{clfv} then results in lower bounds on the photon factor as a function of \RPhaseInput{} and \Rangle{}. 
These lower bounds are shown as contour plots in figure~\ref{fig:current constraints} for each neutrino mass ordering. 
Currently, \mueg{} is the most constraining observable in the largest
part of the \plane{}.
There are only two areas in this plane where \mueg{} gives weaker
constraints than \taueg{} and \taumug{}, corresponding to areas of
approximately zero Yukawa couplings $\flavorY{2}{e}$, $\flavorY{2}{\mu}=0$. 
Hence, the final, absolute constraints on the photon
factor \photonFactor{} (which are independent of \RPhaseInput{} and \Rangle{}) come from $\tau$ decays and are given separately in table~\ref{tab:special point limits}. 

In the mentioned special parameter areas, the observation of \taueg{} or \taumug{} in the planned experiments becomes possible and leads to the regions shown in figure~\ref{fig:optimistic regions}.
The upper limits on \photonFactor{} for these decays is given in  table~\ref{tab:optimistic limits from above}. 
Hence, if \taueg{} or \taumug{} is observed in the future, the parameter space is drastically reduced, while the observation of both \taueg{} and \taumug{} is excluded in the \gls{gnm} by the sensitivity of \mueg{}. 

To disentangle the discussion of the neutrino sector and the scalar
sector, we look at global bounds on \photonFactor{}, which are
independent of the neutrino parameters  \RPhaseInput{} and \Rangle{}.
One option is to consider \emph{absolute} bounds, i.e.\ bounds for
which the complete \plane{} is strictly excluded. 
To increase this absolute bound  an improvement of the
sensitivity in \taueg{} is most important in the near future. 
This situation is shown in figure~\ref{fig:all-pessimist-3}, where the needed branching ratios of \taueg{} and \taumug{} to exclude the particular value of \photonFactor{} for the complete parameter space of the model can be extracted. 
For instance, until the $\br(\taueg)\approx 10^{-11} (10^{-10})$
for \gls{nh}(\gls{ih}) is probed, any other improvement of  \gls{clfv} observables has no effect on the bound on the photon factor \photonFactor{}.
Technically, the absolute bound is driven by the point, where $\flavorY2{\mu}=0$, see table~\ref{tab:special point limits}.
In the numerical evaluation it is helpful to rely on analytical solutions for such points, provided in section~\ref{sub:special-yukawas}. 

A second option is to consider \emph{typical} bounds, i.e.~bounds
on \photonFactor{} for which the essential part of the \plane{} is
excluded.
The absolute and typical bounds are shown in eq.~\eqref{eq:main result} and interpreted in terms of the scalar sector parameter $\lambda_5$ in eq.~\eqref{eq:main result l5}. 

All the presented results are directly applicable to the scoto-seesaw model and complement the current studies of~\cite{Rojas:2018wym, Mandal:2021yph}. 
Additionally, the presented neutrino mass calculation and the parameterization of the Yukawa couplings are more accurate and do not neglect the mixing between the radiative and the seesaw neutrino states at one loop as in~\cite{Rojas:2018wym, Mandal:2021yph}. 
This allows us to cover the parameter space completely.

One cannot apply the absolute bound from the \gls{gnm} directly to the scotogenic model.
In both models, the dependence on the photon factor coincides, but there are more free parameters in the Yukawa sector of the scotogenic model. 
This leads to unrelated analytical solutions in the limiting case.
However, as discussed in section~\ref{sec:lam5}, the typical bounds of eqs.~(\ref{eq:main result},~\ref{eq:main result l5}) are applicable for the typical behavior of the scotogenic model.
\acknowledgments

W.K. was supported in part by the German Research Foundation (DFG) under grant number  STO 876/2-2 and by the National Science Centre (Poland) under the research grant 2020/38/E/ST2/00126.
U.Kh. was supported by the Deutscher Akademischer Austauschdienst (DAAD) under Research Grants --- Doctoral Programmes in Germany, 2019/20 (57440921)
and by the DFG under grant number STO 876/7-1.
This project has received funding from the European Social Fund (project No 09.3.3-LMT-K-712-19-0013) under the grant agreement with the Research Council of Lithuania (LMTLT).

\appendix 	
\section{Parametrization for Yukawa couplings}
\label{app:parametrization}
In this appendix we give the step by step derivation of how we can define the Yukawa couplings unambiguously. 

We start with defining the loop contribution to \oneLoopNuMatrix, eq.~\eqref{eq:mass matrix}, as 
\begin{equation}
W := 
-\Lambda
\loopRotation{2}^*
\bigg(\begin{matrix}
d^2 & i d d^\prime \\
i d d^\prime & -d^{\prime 2}
\end{matrix}\bigg)
\loopRotation{2}^\dagger
=
\loopRotation{2}^{*}
\bigg(\begin{matrix}
0 & 0\\
0 & m_{3}
\end{matrix}\bigg)\loopRotation{2}^\dagger-\text{diag}(\pole{2}\,, \pole{3})
\,,
\label{eq:singular matrix}
\end{equation}
which has determinant zero. 
For convenience we repeat the definition of \loopRotation{2}:
\begin{equation}
\loopRotation{2} = \bigg(\begin{matrix}
R_{22} & -R_{32}^*e^{i\RPhaseOverall} \\
R_{32} & \hphantom{-}R_{22}^*e^{i\RPhaseOverall}
\end{matrix}\bigg)
\,,\quad\text{with}\quad
R_{22}:=\cos\Rangle\ e^{i\RPhaseInput}
\,,\quad
R_{32}:=\sin\Rangle\ e^{i\RPhaseReplaced}
\,.
\end{equation}
Defining 
\begin{equation}
z := R_{22}^{2} + \nuratio \indexvalign R_{32}^{2}
\quad\text{and}\quad
\nuratio := \frac{m_{3}^{\text{pole}}}{m_{2}^{\text{pole}}}
\,,
\label{eq:R relations1}
\end{equation}
we can restrict the components of \loopRotation{2} by the condition, 
that the determinant of the \rhs{} of eq.~\eqref{eq:singular matrix}
vanishes: 
\begin{equation}
\begin{aligned}
  0
=& \,
  \det W
= 
  m_{2}^{\text{pole}} m_{3}^{\text{pole}}
-
  e^{-2i\RPhaseOverall} m_{3}
  ( m_{2}^{\text{pole}} R_{22}^{2} + m_{3}^{\text{pole}} R_{32}^{2} )
\\
=& \,
  m_{2}^{\text{pole}} 
[
  m_{3}^{\text{pole}}
-
  e^{-2i\RPhaseOverall} m_{3}
  ( R_{22}^{2} + \nuratio \indexvalign R_{32}^{2} )
] 
=  
  m_{2}^{\text{pole}} 
[
  m_{3}^{\text{pole}}
-
  e^{-2i\RPhaseOverall} m_{3} z
] 
\,,
\label{eq:R relations2}
\end{aligned}
\end{equation}
giving us a definition of the phase \RPhaseOverall:
\begin{equation}
\frac{e^{2i\RPhaseOverall}}{z} =
\frac{m_{3}}{m_{3}^{\text{pole}}} > 0
\,.
\label{eq:R relations}
\end{equation}

$W$ itself, eq.~\eqref{eq:singular matrix}, produces also additional relations 
between $d$, $d^\prime$ and $\loopRotation{2}_{ij}$: 
\begin{equation}
  [-\loopRotation{2}^{T} W \loopRotation{2}^{}]_{11}
=
  d^2\Lambda 
=
  \pole{2} R_{22}^2 + \pole{3} R_{32}^2
=
  \pole{2}z 
\,.
\label{eq:getting-dis}
\end{equation}
Since $\Lambda$ is real, see eq.~\eqref{eq:mass matrix}, 
we get from eq.~\eqref{eq:getting-dis} 
\begin{equation}
\Lambda / z = \pole{2} / d^{2} > 0 
\,,\quad 
\im\,z=0
\,, 
\label{eq:im z}
\end{equation}
meaning that $z$ has to be real, and additionally, 
has to have the same sign as $\Lambda$. 
From eq.~\eqref{eq:R relations} we get then 
\begin{equation}
  e^{2i\RPhaseOverall} 
=
  \sign(z) 
=
  \sign(\Lambda) 
\,,
\label{eq: 2 RPhaseOverall}
\end{equation}
which allows two values for 
$e^{i\RPhaseOverall}$, depending on the sign of $\Lambda$:
\begin{equation}
  e^{i\RPhaseOverall} 
=
\pm \sqrt{\sign(\Lambda)} 
\,\quad\text{or equivalently}\quad
e^{i\RPhaseOverall}\big|_{\Lambda>0} = \pm 1 
\quad\text{or}\quad
e^{i\RPhaseOverall}\big|_{\Lambda<0} = \pm i
\,.
\label{eq:conditions}
\end{equation}
From our limit on the range of $m_{3}$, eq.~\eqref{eq:consistency}, we get also a limit for $|z| = \pole{3} / m_{3}$:
\begin{equation}
\frac{1}{2} < |z| \leq \nuratio
\,.
\label{eq:z bounds}
\end{equation}
Using above definitions and conditions, eqs.~(\ref{eq:R relations1},~\ref{eq:R relations}), it turns out¨ that we can rewrite 
\begin{equation}
\begin{aligned}
  W_{\rhs}
=& \,
\bigg(\begin{matrix}
- m_{2}^{\text{pole}} + e^{-2i\RPhaseOverall} m_{3} R_{32}^{2}
& - e^{-2i\RPhaseOverall} m_{3} R_{22}^{} R_{32}^{}
\\
- e^{-2i\RPhaseOverall} m_{3} R_{22}^{} R_{32}^{}
& - m_{3}^{\text{pole}} + e^{-2i\RPhaseOverall} m_{3} R_{22}^{2}
\end{matrix}\bigg)
\\
=& \, 
\bigg(\begin{matrix}
- m_{2}^{\text{pole}} + \frac{\nuratio  m_{2}^{\text{pole}}}{z m_{3}} m_{3} R_{32}^{2}
& - \frac{\nuratio  m_{2}^{\text{pole}}}{z m_{3}} m_{3} R_{22}^{} R_{32}^{}
\\
- \frac{\nuratio  m_{2}^{\text{pole}}}{z m_{3}} m_{3} R_{22}^{} R_{32}^{}
& - \nuratio  m_{2}^{\text{pole}} + \frac{\nuratio  m_{2}^{\text{pole}}}{z m_{3}} m_{3} R_{22}^{2}
\end{matrix}\bigg)
\\
=& \, - \frac{m_{2}^{\text{pole}}}{z} 
\bigg(\begin{matrix}
z - \nuratio  R_{32}^{2}
& \nuratio  R_{22}^{} R_{32}^{}
\\
\nuratio  R_{22}^{} R_{32}^{}
& z \nuratio  - \nuratio  R_{22}^{2}
\end{matrix}\bigg)
\\
=& \, - \frac{m_{2}^{\text{pole}}}{z} 
\bigg(\begin{matrix}
R_{22}^{2} + \nuratio \indexvalign R_{32}^{2} - \nuratio  R_{32}^{2}
& \nuratio  R_{22}^{} R_{32}^{}
\\
\nuratio  R_{22}^{} R_{32}^{}
& \nuratio  ( R_{22}^{2} + \nuratio  \indexvalign R_{32}^{2} - R_{22}^{2} )
\end{matrix}\bigg)
\\
=& \, - \frac{m_{2}^{\text{pole}}}{z} 
\bigg(\begin{matrix}
R_{22}^{2} 
& \nuratio  R_{22}^{} R_{32}^{}
\\
\nuratio  R_{22}^{} R_{32}^{}
& \nuratio^{2} {\indexvalign} R_{32}^{2} 
\end{matrix}\bigg)
\,
\label{eq:Wrhs}
\end{aligned}
\end{equation}
as a tensor product:
\begin{equation}
W_{\rhs}
=
-\frac{\pole{2}}{z}\ w\otimes w
\,,\quad
w :=
\left(\begin{array}{c}
R_{22}\\
\nuratio R_{32}
\end{array}\right)
\,.
\label{eq:decomposing singular}
\end{equation}
Motivated by this decomposition we can extend the tensor product definition, 
using eq.~\eqref{eq:GN parameters}, and \loopRotation{2} from eq.~\eqref{eq:four-four-matrices} to express:
\begin{equation}
\massY{2}{}\otimes\massY{2}{}
=
\left(\begin{array}{cc}
0 & 0\\
0 & -\frac{1}{\Lambda} W_{\lhs}
\end{array}\right)
=
\left(\begin{array}{cc}
0 & 0\\
0 & -\frac{1}{\Lambda} W_{\rhs}
\end{array}\right)
=
\left(\begin{array}{cc}
0 & 0\\
0 & \frac{\pole{2}}{z\Lambda}\ w\otimes w
\end{array}\right)
\,.
\end{equation}
This expression motivates a parameterization for the Yukawa coupling $\massY{2}{}$, 
\begin{equation}
\massY{2}{}=
\sign(\Lambda)\,
\sqrt{\frac{\pole{2}}{z\Lambda}}
\
(
0\,, R_{22}\,, \nuratio R_{32}
)
 = 
\niceY{2}{} \loopRotation{3}^\dagger
=
( 0 \,, d \,, i d^{\prime} ) \loopRotation{3}^\dagger
\,,
\label{eq:Y2 mass eigenstates}
\end{equation}
or
\begin{equation}
\begin{aligned}
  d
=& \,
\sign(\Lambda)\,
\sqrt{\frac{\pole{2}}{z\Lambda}}
\,
( R_{22}^{2} + \nuratio  R_{32}^{2} ) 
= 
\frac{\Lambda}{|\Lambda|}
\sqrt{\frac{\pole{2}}{z\Lambda}}
\, z
= 
\frac{\sqrt{\pole{2}z\Lambda}}{|\Lambda|}
> 0
\,,
\\
  i d^{\prime} 
= & \,
  \sign(\Lambda)\,
  \sqrt{\frac{\pole{2}}{z\Lambda}}
\,
  (- R_{22} \indexvalign R_{32}^*
  + \nuratio  R_{32} \indexvalign R_{22}^*
  ) e^{i\RPhaseOverall}
\,,
\end{aligned}
\label{eq:d dprime from Y2}
\end{equation}
which is consistent with the 
$[-\loopRotation{2}^{T} W \loopRotation{2}^{}]_{12}$ 
element of eq.~\eqref{eq:singular matrix} 
\begin{equation}
\begin{aligned}
  i d d^{\prime}\Lambda 
=& \,
  e^{i\RPhaseOverall} (\pole{3} R_{32}\indexvalign R_{22}^*
  - \pole{2} R_{22}\indexvalign R_{32}^*)
\\
=& \,
  \pole{2} e^{i\RPhaseOverall} ( \nuratio  R_{32}\indexvalign R_{22}^*
  - R_{22}\indexvalign R_{32}^*)
\\
=& \,
  \pole{2} \sin\Rangle\cos\Rangle e^{i\RPhaseOverall} 
  ( \nuratio  e^{i ( \RPhaseReplaced - \RPhaseInput )} 
  -  e^{-i ( \RPhaseReplaced - \RPhaseInput )} )
\,.
\label{eq:getting-dprime}
\end{aligned}
\end{equation}
Analogously, we have to rotate the Yukawa coupling to the first Higgs doublet:
\begin{equation}
  \massY{1}{}  
=
  \niceY{1}{} \loopRotation{3}^\dagger 
=
  ( 0 \,, 0 \,, i y ) \loopRotation{3}^\dagger 
=
i e^{-i\RPhaseOverall}\sqrt{\frac{2\pole{3}m_4}{|z|v^2}}\big(0\,, -R_{32}\,, R_{22}\big)
\,,
\label{eq:Y1 mass eigenstates}
\end{equation}
using eq.~(\ref{eq:seesaw relations},~\ref{eq:R relations}).

\section{Determining the minimal parameter space}
\label{app:parameter space}

In this appendix we derive a minimal region of \RPhaseOverall, \Rangle{}, 
\RPhaseInput{}, and \RPhaseReplaced{} that covers the whole parameter space 
to study the unique constraints by \gls{clfv}.
The dominant contribution of all \gls{clfv} processes in our region of 
interest, that is motivated by the approximate $Z_2$ symmetry, 
depends on $\flavorY{2}{i}\flavorY{2}{j}\vphantom{Y}^{*\vphantom{)}}_{\vphantom{i}}$, 
which implies, that parameter points leading to a different 
overall \emph{phase} of \flavorY{2}{} are equivalent.

With the new definitions for \massY{}{} from eq.~\eqref{eq:Y2 mass eigenstates}
\begin{equation*}
\massY{2}{}
=
\sign(\Lambda)\,
\sqrt{\frac{\pole{2}}{z\Lambda}}
\, ( 0\,, R_{22}\,, \nuratio  R_{32} )
=
\sign(\Lambda)\,
\sqrt{\frac{\pole{2}}{z\Lambda}}
\, ( 0\,, \cos\Rangle\ e^{i\RPhaseInput}\,, \nuratio  \sin\Rangle\ e^{i\RPhaseReplaced} )
\,,
\end{equation*}
and eq.~\eqref{eq:Y1 mass eigenstates}
\begin{equation*}
  \massY{1}{}  
=
  \niceY{1}{} \loopRotation{3}^\dagger 
=
  ( 0 \,, 0 \,, i y ) \loopRotation{3}^\dagger 
=
i e^{-i\RPhaseOverall}\sqrt{\frac{2\pole{3}m_4}{|z|v^2}}
\big(0\,, -\sin\Rangle\ e^{i\RPhaseReplaced}\,, \cos\Rangle\ e^{i\RPhaseInput}\big)
\,,
\end{equation*}
we see the following relation
\begin{equation}
	\massY{}{}\left(\Rangle \pm \pi \right) = -\massY{}{}\left(\Rangle \right)
	\quad
	\Rightarrow
	\quad
	\Rangle\in\big(-\tfrac{\pi}{2},\tfrac{\pi}{2}\big]
	\,,
\end{equation}
that reduces the required region of \Rangle{}, as the minus signs are just 
a different phase, that does not influence the \gls{clfv}.
In the same way, we can pick in eq.~\eqref{eq:conditions} the plus sign 
before the square root, as the different phase does not influence the \gls{clfv}.

The restrictions for \Rangle{} and \RPhaseInput{} come from 
solving eq.~\eqref{eq:im z} for \RPhaseReplaced{}:
\begin{equation}
\begin{aligned}
  0 
= 
  \im\,z
= 
  \im[ R_{22}^{2} + \nuratio R_{32}^{2} ]
=& \,
  \cos^{2}\Rangle\ \sin (2\RPhaseInput)
+ \nuratio \sin^{2}\Rangle\ \sin(2\RPhaseReplaced)
\\
=& \,
  \cos^{2}\Rangle\ \sin (2\RPhaseInput)
+ \nuratio \sin^{2}\Rangle\ 
\frac{1}{2i}
[ e^{2i\RPhaseReplaced} - e^{-2i\RPhaseReplaced} ]
\end{aligned}
\label{eq:im z new}
\end{equation}
gives two possible distinct solutions for \RPhaseReplaced{} as functions of 
\RPhaseInput{}, as $e^{i\RPhaseReplaced}$ 
is needed for the Yukawas, but the equation determines only the square, 
$e^{2i\RPhaseReplaced}$: 
\begin{equation}
  0 
= \,
  e^{2i\RPhaseReplaced}
+ 2 i \frac{\sin (2\RPhaseInput)}{\nuratio \tan^{2}\Rangle}
- e^{-2i\RPhaseReplaced}
\,,
\end{equation}
or
\begin{equation}
  ( e^{2i\RPhaseReplaced} )_{1, 2}^{}
=
- i \frac{\sin (2\RPhaseInput)}{\nuratio \tan^{2}\Rangle}
\pm \sqrt{1 - \frac{\sin^{2} (2\RPhaseInput)}{\nuratio^{2} \tan^{4}\Rangle}}
\,,
\label{eq:e 2i RPhaseReplaced}
\end{equation}
giving immediately
\begin{equation}
  [ \sin(2\RPhaseReplaced) ]_{1}^{}
=
  [ \sin(2\RPhaseReplaced) ]_{2}^{}
=
- \frac{\sin (2\RPhaseInput)}{\nuratio \tan^{2}\Rangle}
\label{eq: sin 2 RPhaseReplaced}
\end{equation}
and
\begin{equation}
  [ \cos(2\RPhaseReplaced) ]_{1,2}^{}
=
\pm \sqrt{1 - \frac{\sin^{2} (2\RPhaseInput)}{\nuratio^{2} \tan^{4}\Rangle}}
\,.
\label{eq: cos 2 RPhaseReplaced}
\end{equation}
One can directly see, that $( e^{2i\RPhaseReplaced} )_{2} 
= [-( e^{2i\RPhaseReplaced} )_{1} ]^{-1}$.
Additionally, eqs.~(\ref{eq: sin 2 RPhaseReplaced},~\ref{eq: cos 2 RPhaseReplaced}) restrict \Rangle{} and \RPhaseInput{} by
\begin{equation}
  \nuratio \tan^{2}\Rangle > | \sin (2\RPhaseInput) |
\, ,
\label{eq: excluded region}
\end{equation}
which is displayed as the white ``disks'' in figure~\ref{fig:omega-r-region}.
In these white ``disks'' there exists no solution to consistency conditions 
imposed by eq.~\eqref{eq:singular matrix}.

We take as the regular solution, inspired by eq.~\eqref{eq: sin 2 RPhaseReplaced}:
\begin{equation}
\positiveSolution :=
-\frac{1}{2}\arcsin\left( \frac{\sin(2\RPhaseInput{})}{\nuratio \tan^{2}\Rangle{}} \right)
\,,\quad
\positiveSolution(\Rangle=0) := 0
\,,
\label{eq:w32+}
\end{equation}
which lies by construction in the range $|\RPhaseReplaced|<\pi/4$, giving 
$\cos(2\RPhaseReplaced) \ge 0$. To reach the second choice, 
$\cos(2\RPhaseReplaced)<0$, we take: 
\begin{equation}
\negativeSolution := -\frac{\pi}{2} \sign(\RPhaseInput) -\positiveSolution
\,.
\label{eq:w32-}
\end{equation}
Both \positiveSolution{} and \negativeSolution{} lead to regions with 
positive and negative values for $z$: 
\begin{equation}
\begin{aligned}
  z_{+} 
:= & \,
  \operatorname{Re}\,[z (\positiveSolution)]
= 
  \operatorname{Re}[ R_{22}^{2} + \nuratio R_{32}^{2} ]
= \,
  \cos^{2}\Rangle\ \cos (2\RPhaseInput)
+ \nuratio \sin^{2}\Rangle\ \cos(2\positiveSolution)
\\
=& \,
  \cos^{2}\Rangle\ 
  \left[ \cos (2\RPhaseInput)
  + \nuratio \tan^{2}\Rangle\ 
    \sqrt{1 - \frac{\sin^{2} (2\RPhaseInput)}{\nuratio^{2} \tan^{4}\Rangle}} 
  \right]
\,.
\end{aligned}
\label{eq:re z+}
\end{equation}
If 
\begin{equation}
\begin{aligned}
& \nuratio^2 \tan^{4}\Rangle\ 
    \left[ 1 - \frac{\sin^{2} (2\RPhaseInput)}{\nuratio^{2} \tan^{4}\Rangle}
    \right] 
\ge 
   \cos^{2} (2\RPhaseInput)
\\
\Leftrightarrow \qquad
&
  \nuratio^2 \tan^{4}\Rangle\ 
\ge 
   \cos^{2} (2\RPhaseInput)
+ \nuratio^2 \tan^{4}\Rangle\ 
  \frac{\sin^{2} (2\RPhaseInput)}{\nuratio^{2} \tan^{4}\Rangle}
= 
  1
\\
\Leftrightarrow \qquad
&
  \nuratio \tan^{2}\Rangle\ 
\ge 
  1
\,,
\end{aligned}
\label{eq:condition z+}
\end{equation}
$z_{+} > 0$ for any value of \RPhaseInput{}. But if 
\begin{equation}
  \nuratio \tan^{2}\Rangle \le 1
\label{eq:condition z-}
\end{equation}
$z_{+}$ can become negative for $|\RPhaseInput{}| > \frac{\pi}{4}$. 

A similar distinction happens to
\begin{equation}
\begin{aligned}
  z_{-} 
:= & \,
  \operatorname{Re}\,[z (\negativeSolution)]
= 
  \operatorname{Re}[ R_{22}^{2} + \nuratio R_{32}^{2} ]
= \,
  \cos^{2}\Rangle\ \cos (2\RPhaseInput)
+ \nuratio \sin^{2}\Rangle\ \cos(2\negativeSolution)
\\
=& \,
  \cos^{2}\Rangle\ \cos (2\RPhaseInput)
+ \nuratio \sin^{2}\Rangle\ \cos(\mp \pi -2\positiveSolution)
\\
=& \,
  \cos^{2}\Rangle\ \cos (2\RPhaseInput)
- \nuratio \sin^{2}\Rangle\ \cos(-2\positiveSolution)
\\
=& \,
  \cos^{2}\Rangle\ 
  \left[ 
    \cos (2\RPhaseInput)
  - \nuratio \tan^{2}\Rangle\ 
    \sqrt{1 - \frac{\sin^{2} (2\RPhaseInput)}{\nuratio^{2} \tan^{4}\Rangle}}
  \right]
\,.
\end{aligned}
\label{eq:re z-}
\end{equation}
For eq.~\eqref{eq:condition z+} the negative term always dominates and 
$z_{-} \le 0$ for any \RPhaseInput{}. But if eq.~\eqref{eq:condition z-}, 
the first term dominates and $z_{-} \ge 0$ if $|\RPhaseInput{}| < \frac{\pi}{4}$.

Since $z$ and $\Lambda$ have the same sign, eq.~\eqref{eq:im z}, 
the sign of $\Lambda$ determines also, how the choice of \RPhaseInput, 
effects the possible solutions of eq.~\eqref{eq:im z} or eq.~\eqref{eq:im z new},
i.e. \positiveSolution{} or \negativeSolution{}, as the sign of $z$ is an 
additional requirement for the solution: 
\begin{itemize}
\item in the case of a positive sign of $\Lambda$, we have to choose also a 
positive sign for $z$:
\begin{itemize}
\item $z_{+}$ in the region eq.~\eqref{eq:condition z+} and the region
eq.~\eqref{eq:condition z-} and $|\RPhaseInput{}| < \frac{\pi}{4}$
\item $z_{-}$ in the region 
eq.~\eqref{eq:condition z-} and $|\RPhaseInput{}| > \frac{\pi}{4}$
\end{itemize}
\item in the case of a negative sign of $\Lambda$, we have to choose a 
negative sign for $z$:
\begin{itemize}
\item $z_{+}$ in the region 
eq.~\eqref{eq:condition z-} and $|\RPhaseInput{}| > \frac{\pi}{4}$
\item $z_{-}$ in the region eq.~\eqref{eq:condition z+} and the region
eq.~\eqref{eq:condition z-} and $|\RPhaseInput{}| < \frac{\pi}{4}$
\end{itemize}
\end{itemize}
Summarizing, we get the possible Yukawa couplings:
\begin{align}
&  \massY{}{}(\Lambda , \enspace \nuratio \tan^{2}\Rangle\, <  |\sin(2\RPhaseInput)| ) 
&:= & \, 
  \text{not defined}
\\[5pt]
&  \massY{}{}(\Lambda > 0, \nuratio  > \cot^{2}\Rangle\, ) 
&:= & \, 
  \massY{}{}(\Rangle, \RPhaseInput; \positiveSolution{} ) 
\\
&  \massY{}{}(\Lambda > 0, \nuratio  < \cot^{2}\Rangle\, , |\RPhaseInput| < \pi/4 ) 
&:= & \, 
  \massY{}{}(\Rangle, \RPhaseInput; \positiveSolution{} ) 
\\
&  \massY{}{}(\Lambda > 0, \nuratio  < \cot^{2}\Rangle\, , |\RPhaseInput| > \pi/4 ) 
&:= & \, 
  \massY{}{}(\Rangle, \RPhaseInput; \negativeSolution{} ) 
\\[5pt]
&  \massY{}{}(\Lambda < 0, \nuratio  < \cot^{2}\Rangle\, , |\RPhaseInput| > \pi/4 ) 
&:= & \, 
  \massY{}{}(\Rangle, \RPhaseInput; \positiveSolution{} ) 
\\
&  \massY{}{}(\Lambda < 0, \nuratio  > \cot^{2}\Rangle\, ) 
&:= & \, 
  \massY{}{}(\Rangle, \RPhaseInput; \negativeSolution{} ) 
\\
&  \massY{}{}(\Lambda < 0, \nuratio  < \cot^{2}\Rangle\, , |\RPhaseInput| < \pi/4 ) 
&:= & \, 
  \massY{}{}(\Rangle, \RPhaseInput; \negativeSolution{} ) 
\,.
\label{eq:fully-defined-yukawas}
\end{align}
We see, that the ares where one has to use the ``other'' solution, 
i.e. \negativeSolution{} instead of \positiveSolution{}, or 
$\Lambda > 0$ compared to $\Lambda < 0$, are complementary: we cover the 
whole parameter space in all cases. 

\section{Relations between Yukawa couplings of different parameter points}
\label{app:Yukawa relations}

It turns out, that there is a relation between the parameters of 
\loopRotation{2}, that gives the same Yukawa couplings for both solutions, 
\positiveSolution{} and \negativeSolution{}. 
We also notice that the conditions for the allowed areas in the \plane{} 
depend only on the sizes of $|\Rangle|$ and $|\RPhaseInput|$.

The first observation needed is the relation between 
$\positiveSolution{}[\RPhaseInput{}]$, eq.~\eqref{eq:w32+}, and 
$\negativeSolution{}[\widetilde{\RPhaseInput}]$, eq.~\eqref{eq:w32-}, 
for a shifted 
$\widetilde{\RPhaseInput} = \RPhaseInput - \frac{\pi}{2} \sign(\RPhaseInput)$. 
Please note, that this shift flips between the regions 
$|\RPhaseInput|<\frac{\pi}{4}$ and $|\RPhaseInput|>\frac{\pi}{4}$. 
\begin{equation}
\begin{aligned}
  \negativeSolution{}[\widetilde{\RPhaseInput}]
=& \enspace
  \negativeSolution[\RPhaseInput - \frac{\pi}{2} \sign(\RPhaseInput) ] 
\\
=& 
- \frac{\pi}{2} \sign(\RPhaseInput) 
+ \frac{1}{2}\arcsin\left( 
    \frac{\sin\left(2[\RPhaseInput - \frac{\pi}{2} \sign(\RPhaseInput) ]\right)}
         {\nuratio \tan^{2}\Rangle{}} \right)
\\
=& 
- \frac{\pi}{2} \sign(\RPhaseInput) 
+ \frac{1}{2}\arcsin\left( 
    \frac{\sin\left(2\RPhaseInput \mp \pi \right)}
         {\nuratio \tan^{2}\Rangle{}} \right)
\\
=& 
- \frac{\pi}{2} \sign(\RPhaseInput) 
+ \frac{1}{2}\arcsin\left(
    \frac{- \sin\left(2\RPhaseInput\right)}
         {\nuratio \tan^{2}\Rangle{}} \right)
\\
=& 
- \frac{\pi}{2} \sign(\RPhaseInput) 
- \frac{1}{2}\arcsin\left(
    \frac{\sin\left(2\RPhaseInput\right)}
         {\nuratio \tan^{2}\Rangle{}} \right)
\\
=& 
- \frac{\pi}{2} \sign(\RPhaseInput) 
+ \positiveSolution{}[\RPhaseInput{}]
\,.
\end{aligned}
\end{equation}
The second relation is the one between $z_{-}(\widetilde{\RPhaseInput})$, 
eq.~\eqref{eq:re z-}, and $z_{+}(\RPhaseInput)$, eq.~\eqref{eq:re z+}:
\begin{equation}
\begin{aligned}
  z_{-} 
= & \,
  \cos^{2}\Rangle\ \cos (2\widetilde{\RPhaseInput})
+ \nuratio \sin^{2}\Rangle\ \cos(2\negativeSolution[\widetilde{\RPhaseInput}])
\\
=& \,
  \cos^{2}\Rangle\ \cos (2[\RPhaseInput - \sfrac{\pi}{2} \sign(\RPhaseInput)])
+ \nuratio \sin^{2}\Rangle\ 
  \cos(2[ \positiveSolution{}[\RPhaseInput{}] - \sfrac{\pi}{2} \sign(\RPhaseInput)])
\\
=& \,
  \cos^{2}\Rangle\ \cos (2\RPhaseInput \mp \pi )
+ \nuratio \sin^{2}\Rangle\ 
  \cos(2 \positiveSolution{}[\RPhaseInput{}] \mp \pi )
\\
=& \,
- \cos^{2}\Rangle\ \cos (2\RPhaseInput )
- \nuratio \sin^{2}\Rangle\ 
  \cos(2 \positiveSolution{}[\RPhaseInput{}] )
\, = \, 
- z_{+}(\RPhaseInput)
\,.
\end{aligned}
\end{equation}
With $\sign(\Lambda) = \sign(z) = +1$
this gives now  
\begin{equation}
\begin{aligned}
&  \massY{1}{}(\Lambda > 0, \nuratio  < \cot^{2}\Rangle\,
              , |\widetilde{\RPhaseInput}| > \pi/4 )
= \massY{1}{}( \Rangle, \widetilde{\RPhaseInput};
              \negativeSolution[\widetilde{\RPhaseInput}] )
\\
&=
i \sqrt{\sign(z_{-})}\sqrt{\frac{2\pole{3}m_4}{|z_{-}[\widetilde{\RPhaseInput}]|v^2}}
\big(0
, -\sin\Rangle\ e^{i\negativeSolution[\widetilde{\RPhaseInput}]}
, \cos\Rangle\ e^{i\widetilde{\RPhaseInput}}
\big)
\\
&=
i \sqrt{\sign(-z_{+})}\sqrt{\frac{2\pole{3}m_4}{|-z_{+}[\RPhaseInput]|v^2}}
\big(0
, -\sin\Rangle\ e^{i\positiveSolution[\RPhaseInput] - \frac{i\pi}{2} \sign(\RPhaseInput)}
, \cos\Rangle\ e^{i\RPhaseInput - \frac{i\pi}{2} \sign(\RPhaseInput)}
\big)
\\
&=
i \sqrt{-\sign(z_{+})} \, e^{- \frac{i\pi}{2} \sign(\RPhaseInput)} \, 
\sqrt{\frac{2\pole{3}m_4}{|z_{+}[\RPhaseInput]|v^2}}
\big(0
, -\sin\Rangle\ e^{i\positiveSolution[\RPhaseInput]}
, \cos\Rangle\ e^{i\RPhaseInput}
\big)
\\
&=
\sqrt{- e^{- i\pi \sign(\RPhaseInput)}} \, 
\massY{1}{}(\Rangle, \RPhaseInput; \positiveSolution[\RPhaseInput] )
\\
&=
\massY{1}{}(\Rangle, \RPhaseInput; \positiveSolution[\RPhaseInput] )
\, = \, 
  \massY{1}{}(\Lambda > 0, \nuratio  < \cot^{2}\Rangle\,
              , |\RPhaseInput| < \pi/4 )
\,,
\end{aligned}
\label{eq:Y1 m-}
\end{equation}
or with $\sign(\Lambda) = \sign(z) = -1$ we get 
\begin{equation}
\begin{aligned}
&  \massY{1}{}(\Lambda < 0, \nuratio  < \cot^{2}\Rangle\,
              , |\widetilde{\RPhaseInput}| < \pi/4 )
= \massY{1}{}( \Rangle, \widetilde{\RPhaseInput};
              \negativeSolution[\widetilde{\RPhaseInput}] )
\\
&=
i \sqrt{\sign(z_{-})}\sqrt{\frac{2\pole{3}m_4}{|z_{-}[\widetilde{\RPhaseInput}]|v^2}}
\big(0
, -\sin\Rangle\ e^{i\negativeSolution[\widetilde{\RPhaseInput}]}
, \cos\Rangle\ e^{i\widetilde{\RPhaseInput}}
\big)
\\
&=
i \sqrt{\sign(-z_{+})}\sqrt{\frac{2\pole{3}m_4}{|-z_{+}[\RPhaseInput]|v^2}}
\big(0
, -\sin\Rangle\ e^{i\positiveSolution[\RPhaseInput] - \frac{i\pi}{2} \sign(\RPhaseInput)}
, \cos\Rangle\ e^{i\RPhaseInput - \frac{i\pi}{2} \sign(\RPhaseInput)}
\big)
\\
&=
i \sqrt{-\sign(z_{+})} \, e^{- \frac{i\pi}{2} \sign(\RPhaseInput)} \, 
\sqrt{\frac{2\pole{3}m_4}{|z_{+}[\RPhaseInput]|v^2}}
\big(0
, -\sin\Rangle\ e^{i\positiveSolution[\RPhaseInput]}
, \cos\Rangle\ e^{i\RPhaseInput}
\big)
\\
&=
\sqrt{- e^{- i\pi \sign(\RPhaseInput)}} \, 
\massY{1}{}(\Rangle, \RPhaseInput; \positiveSolution[\RPhaseInput] )
\\
&=
\massY{1}{}(\Rangle, \RPhaseInput; \positiveSolution[\RPhaseInput] )
\, = \, 
  \massY{1}{}(\Lambda < 0, \nuratio  < \cot^{2}\Rangle\,
              , |\RPhaseInput| > \pi/4 )
\,,
\end{aligned}
\label{eq:Y1 m-2}
\end{equation}
but now for the opposite sign of $\Lambda$.

In the same way we get for any sign of $\Lambda$ 
\begin{equation}
\begin{aligned}
& \massY{2}{}(\Rangle, \widetilde{\RPhaseInput}; \negativeSolution[\widetilde{\RPhaseInput}] ) 
\\
&=
\sign(\Lambda)\,
\sqrt{\frac{\pole{2}}{z_{-}[\widetilde{\RPhaseInput}]\Lambda}}
\, ( 0
, \cos\Rangle\ e^{i\widetilde{\RPhaseInput}}
, \nuratio  \sin\Rangle\ e^{i\negativeSolution[\widetilde{\RPhaseInput}]}
)
\\
&=
\sign(\Lambda)\,
\sqrt{\frac{\pole{2}}{-z_{+}[\RPhaseInput]\Lambda}}
\, ( 0
, \cos\Rangle\ e^{i\RPhaseInput - \frac{i\pi}{2} \sign(\RPhaseInput)}
, \nuratio  \sin\Rangle\
  e^{i\positiveSolution[\RPhaseInput] - \frac{i\pi}{2} \sign(\RPhaseInput)}
)
\\
&=
\sign(\Lambda)\, e^{- \frac{i\pi}{2} \sign(\RPhaseInput)} \sqrt{-1} \,
\sqrt{\frac{\pole{2}}{z_{+}[\RPhaseInput]\Lambda}}
\, ( 0
, \cos\Rangle\ e^{i\RPhaseInput}
, \nuratio  \sin\Rangle\ e^{i\positiveSolution[\RPhaseInput]}
)
\\
&=
\sqrt{- e^{- i\pi \sign(\RPhaseInput)}} \, 
\massY{2}{}(\Rangle, \RPhaseInput; \positiveSolution[\RPhaseInput] )
\, 
=
\, 
\massY{2}{}(\Rangle, \RPhaseInput; \positiveSolution[\RPhaseInput] )
\,,
\end{aligned}
\label{eq:Y2 m-}
\end{equation}
which allows us to cover all values of Yukawa couplings with sweeping over 
the \plane{} with only using \positiveSolution{} for the definition of the
Yukawa couplings, as done in eqs.~(\ref{eq:Y1 mass eigenstates+},~\ref{eq:Y2 mass eigenstates+}).
\section{Numerical values for Yukawas on the \plane{}}\label{app:Y=0 numbers} 

Here we list the numetical values of the special parameter points in the \plane{}. 
For $\flavorY{2}{f}=0$, using the numerical values from eq.~\eqref{eq:neutrino data} for \gls{nh} and putting them into eq.~\eqref{eq:Y2 zero condition} we get:
\begin{eqnarray}
& \flavorY{2}{e}=0 & \Rightarrow  (r, \omega_{22} ) = (0.558002, 1.21395)\,,  \\
& \flavorY{2}{\mu}=0 & \Rightarrow  (r, \omega_{22} ) =  (-0.130718, 0.00977035)\,,   \\
& \flavorY{2}{\tau}=0 & \Rightarrow (r, \omega_{22} ) =  (0.155322,-0.0112205 )\,,   
\end{eqnarray}
and for \gls{ih}: 
\begin{eqnarray}
& \flavorY{2}{e}=0 & \Rightarrow  (r, \omega_{22} ) =  (-0.973831,0 )\,,    \\
& \flavorY{2}{\mu}=0 & \Rightarrow  (r, \omega_{22} ) = (0.559934,-0.0992133 )\,,   \\
& \flavorY{2}{\tau}=0 & \Rightarrow (r, \omega_{22} ) = (0.617427,0.0919401) \,.  
\end{eqnarray}

\bibliographystyle{jhep.bst}
\bibliography{bibliography}
\end{document}